\documentstyle[12pt,epsfig]{article}
\newcommand{\be}{\begin{equation}}
\newcommand{\ee}{\end{equation}}
\newcommand{\bea}{\begin{eqnarray}}
\newcommand{\eea}{\end{eqnarray}}

\def\la{\mathrel{\mathpalette\fun <}}

\def\fun#1#2{\lower3.6pt\vbox{\baselineskip0pt\lineskip.9pt
\ialign{$\mathsurround=0pt#1\hfil##\hfil$\crcr#2\crcr\sim\crcr}}}

\begin{document}

\title{Quark-gluonium content of the
scalar-isoscalar states  $f_0(980)$, $f_0(1300)$, $f_0(1500)$,
$f_0(1750)$, $f_0(1420\;^{+150}_{-\; 70})$ from hadronic decays }
\author{V.V. Anisovich, V.A. Nikonov and A.V. Sarantsev}
\date{}
\maketitle

\begin{abstract}

On the basis of the decay couplings
$f_0 \to \pi\pi$, $K\bar K$, $\eta\eta$, $\eta\eta'$,
which had been  found before, in the study of
analytical
$(IJ^{PC}=00^{++})$-amplitude in the mass range 450-1900 MeV,
we analyse the quark-gluonium content of  resonances
$f_0(980)$, $f_0(1300)$, $f_0(1500)$,
$f_0(1750)$ and the broad state $f_0(1420\;^{+\;150}_{-70})$.
The $K$-matrix technique used in the analysis
makes it possible to evaluate the quark-gluonium content both for the
states with switched-off decay channels (bare states, $f^{bare}_0$) and
the real resonances. We observe significant change of the
quark-gluonium composition in the evolution from  bare states to real
resonances, that is due to the mixing of states in the transitions
$f_0(m_1)\to real\; mesons\to f_0(m_2)$  responsible for
the decay processes as well. For the $f_0(980)$, the analysis
confirmed the
 dominance of  $q\bar q$ component thus proving the $n\bar n/s\bar s$
composition found in the study of the radiative decays. For the mesons
$f_0(1300)$, $f_0(1500)$ and $f_0(1750)$, the hadronic decays do not
allow one to
determine uniquely
the $n\bar n$, $s\bar s$ and gluonium  components
providing relative pecentage
only. The
analysis shows that the broad state  $f_0(1420\;^{+\;150}_{-70})$ can
mix with the flavour singlet $q\bar q$ component only, that is
consistent with gluonium origin of the broad resonance.

\end{abstract}

\section{Introduction}

The present paper continues the investigation of
$(IJ^{PC}=00^{++})$ resonances started in \cite{YF,width}: we analyse
the $K$-matrix solution in which  scalar glueball
located near 1600 MeV appears, before
mixing with neighbouring $q \bar q$ states due to decay
prosesses.

In \cite{YF},
on the basis of experimental data of GAMS group \cite{GAMS}, Crystal
Barrel Collaboration \cite{CrBar} and BNL group \cite{BNL},
the $K$-matrix solution has been found for the waves $00^{++},
10^{++},02^{++},12^{++}$ covering the mass range
450--1900 MeV. Also  masses and total widths of resonances
have been determined for these waves. The following
states have been seen in the scalar-isoscalar sector,
\be
00^{++}: \qquad
f_0(980),\; f_0(1300),\; f_0(1500),\; f_0(1420^{+150}_{-\;70}),\;
f_0(1750)\ .
\ee
In \cite{PDG}, the resonances $f_0(1300)$ and  $f_0(1750)$ are referred
to as $f_0(1370)$ and  $f_0(1710)$.
The broad state $f_0(1420^{+150}_{-\; 70})$ is not included into the
compilation \cite{PDG}; the broad state is denoted in \cite{YF} as
$f_0(1530^{+\;90}_{-250})$ that represent the mean value for three
solutions found in the K-matrix analysis; here we discuss the
solution only where primary glueball is located near 1600 MeV;
in this way, we use the mass of  broad state found in this
solution.

For the scalar-isovector sector, the analysis \cite{YF} points
to the presence of the following resonances in the spectra:
\be
10^{++}: \qquad a_0(980),\;  a_0(1520)\ .
\ee
In the compilation \cite{PDG} the state $a_0(1520)$ is denoted as
$a_0(1450)$.

As to tensor mesons, the following states are seen:
\bea
&&12^{++}: \qquad a_2(1320),\; a_2(1660),\nonumber \\
&&02^{++}: \qquad f_2(1270),\; f_2(1525)\ .
\eea
Although in the analysis \cite{YF} the CERN-M\"unich group data
\cite{C-M} were not included into fiting procedure directy, these data
had been fitted in previous papers \cite{km,km1900}, and it was
specially controlled that  solutions found in \cite{YF} and
\cite{km,km1900} for the $\pi\pi$ channel are in a good agreement with
each other.

For the states shown in (1), (2) and (3), the $K$-matrix
poles and $K$-matrix couplings to channels $\pi\pi,K\bar K,
\eta\eta,\eta\eta',\pi\pi\pi\pi$ have been
found in \cite{YF}. The
$K$-matrix poles are not the amplitude poles, these latter
corresponding to physical resonances, but when the decays are switched
off, the resonance poles turn into the $K$-matrix ones. In the states
related to the $K$-matrix poles there is no cloud of real mesons which
are due to the decay processes, that was the reason to name them "bare
states". The $K$-matrix couplings for transitions $bare \; state \to
\pi\pi,K\bar K, \eta\eta,\eta\eta'$ found in \cite{YF} confirm the
$q\bar q$ nonet classification of  bare states suggested
in \cite{Al-An,UFN}.

Still, the $K$-matrix analysis \cite{YF} does not supply us with
 partial widths of the
resonances directly. To determine couplings for the  transitions
$resonance \to mesons$, auxillary calculations should
be performed to find out residues of the amplitude
poles.
Calculations of the residues have been carried out in \cite{width} for
scalar-isoscalar sector, that gave us the values
of partial widths for the resonances $f_0(980)$, $f_0(1300)$,
$f_0(1500)$, $f_0(1750)$ and broad state
$f_0(1420\;^{+\;150}_{-70})$ decaying into the channels $\pi\pi,
\pi\pi\pi\pi,K\bar K,\eta\eta,\eta\eta'$.

In this paper we use the decay couplings for the reactions
$f_0\to \pi\pi,K\bar K,\eta\eta,\eta\eta'$ as a basis for the
analysis of quark-gluonium content of scalar-isoscalar
resonances $f_0(980)$, $f_0(1300)$, $f_0(1500)$,
$f_0(1750)$, $f_0(1420\;^{+\;150}_{-70})$,
supposing for these states a three-component structure: $n\bar n=
(u\bar u+d\bar d)/\sqrt 2 , \; s\bar s,\; gluonium$. We demonstrate
that  hadronic decays do not determine the weight of all components
providing the correlation between them only.

The knowledge of coupling constants of bare states $f_0^{bare} \to
\pi\pi,K\bar K,\eta\eta,\eta\eta'$ makes it possible
to trace the evolution of the $q\bar q$ and gluonium
components in $f_0$-mesons
by switching on/off the decay channels, thus establishing
constraints for
these components in a studied resonance.

The paper is organized as follows.

Section 2, being introductive, gives general picture of resonances in
the scalar-isoscalar sector in the mass range up to 2300 MeV and
presents $q\bar q$ classification of the $00^{++}$ states. On the basis
of this classification we analyse  the K-matrix solution
with a primary glueball located near 1600 MeV.
Correspondingly, we explore  parameters found for the broad state in
this solution, $f_0(1420^{+150}_{-\; 70})$.

In Section 3 we analyse
quark-gluonium content of the resonances
$f_0(980)$, $f_0(1300)$, $f_0(1500)$,
$f_0(1750)$ and broad state $f_0(1420\;^{+150}_{-\;70})$
basing on the rules of  quark combinatorics for the
$f_0$ states \cite{combinatorics}.

In Section 4, using the $K$-matrix representation for the
$00^{++}$ amplitude, we study the evolution of
the quark-gluonium content of resonances $f_0(980)$,
$f_0(1300)$, $f_0(1500)$, $f_0(1750)$,
$f_0(1420\;^{+150}_{-\;70})$ by varying
 gradually  the strength of the decay channels.

Concluding, we state that our analysis based on the study
of the hadronic decays only is not able to fix
the $q\bar q$/gluonium content of $f_0$ mesons unambiguously.
For qualitative estimate of  $q\bar q$ and gluonium
components, one needs to incorporate additional
information into analysis,
such as partial widths of the $f_0$-meson produced
in $\gamma\gamma$ collisions, rates of radiative decays with the
$f_0$-meson production, or the $f_0$ production ratios in the
decays of heavy mesons.

\section{ Resonances in scalar-isoscalar sector}

The $K$-matrix analysis \cite{YF} performed for
the scalar-isoscalar sector at
450--1900 MeV allows us to
reconstruct analytical form of the amplitude in the region shown in
Fig.  1 by dashed line, since the threshold singularities of the
$00^{++}$ amplitude related to channels  $\pi\pi$, $\pi\pi\pi\pi$,
$K\bar K$, $\eta\eta$, $\eta\eta'$ are correctly taken into account.
The amplitude poles which correspond to the resonances  (1),
 broad state $f_0(1420\;^{+150}_{-\;70})$ included,
are located just in the
area where analytical structure of the amplitude $00^{++}$ is
restored.

Below mass scale of the $K$-matrix analysis \cite{YF} there is a
pole related to the light $\sigma$-meson: in Fig. 1 its position,
$M=(430-i320)$ MeV, is
shown in  accordance with the results of the dispersion relation
$N/D$-analysis \cite{N/D} (the mass region allowed by this analysis is
also shown in Fig. 1).

The pole related to the light $\sigma$-meson, with the mass $M \simeq
450$ MeV, has been obtained in a number of papers, see \cite{PDG} for
details. In the decay $D^+ \to \pi^+\pi^+\pi^-$ the
$\sigma$-meson mass was  found  to be $M=480\pm 40$ MeV \cite{D+};
corresponding pole location is also shown in Fig. 1 by crossed bars.

Above the mass region of the $K$-matrix analysis,
there are resonances $f_0(2030)$, $ f_0(2100)$,
$f_0(2340)$
\cite{RAL,Systematics}.

\subsection{ Classification of scalar bare states}

In \cite{Al-An,UFN}, in terms of bare states, the quark-gluonium
classification of scalar particles has been suggested.
The results of the K-matrix analysis \cite{YF} support this
classification.

The bare state being a member of a $q\bar q$ nonet imposes rigid
restrictions on the $K$-matrix parameters.
The $q\bar q$ nonet of scalars consists of
two scalar-isoscalar states,
$f_0^{bare}(1)$ and $f_0^{bare}(2)$,
 scalar-isovector meson $a_0^{bare}$ and
scalar kaon $K^{bare}_0$. In the leading terms of the
$1/N $-expansion \cite{t'Hooft}, the decays of these four states
into two pseudoscalar mesons
are determined by three parameters only, which are the common
coupling constant
$g$, suppression parameter $\lambda$ for strange quark production
(in the limit of precise $SU(3)_{flavour}$ symmetry
$\lambda=1$) and mixing angle $\varphi$ for $n\bar n=(u\bar u+d\bar
d)/\sqrt{2}$ and $s\bar s$ components in $f_0^{bare}$:
\be
n\bar n \cos \varphi+s\bar s \sin \varphi\ .
\ee
The mixing angle defines the
scalar-isoscalar nonet partners $f_0^{bare}(1)$ and $f_0^{bare}(2)$:
\be
 \varphi(1)- \varphi(2)=90^\circ\ .
\ee
The restrictions imposed on coupling constants allow one to fix
unambigously the basic scalar nonet:
\be
1^3P_0q\bar q:
f_0^{bare}(720\pm 100),\; a_0^{bare}(960\pm 30),\;
K_0^{bare}(1220^{+\; 50}_{-150}),\;f_0^{bare}(1260\pm 30)\ ,
\label{nonet}
\ee
as well as mixing angle for $f_0^{bare}(720)$ and $f_0^{bare}(1260)$:
\be
 \varphi[f_0^{bare}(720)]=-70^{\circ}\; ^{+\; 5^\circ}_{-10^\circ}\ .
\ee
The nonet $1^3P_0q\bar q$ in the form of (\ref{nonet}) has been
suggested in \cite{Al-An}, where the $K$-matrix re-analysis of the
$K\pi$ data \cite{Kpi} has been carried out (bare states and
their couplings for the
$00^{++}$ and $10^{++}$ waves have been found before, in
\cite{km,km1900}).

To establish the nonet of the first radial excitations,  $2^3P_0q\bar
q$, appeared to be more difficult
problem. The $K$-matrix analysis
\cite{YF} gives two scalar-isoscalar states in the region 1200--1650
MeV, their decay couplings satisfying the requirements
imosed for glueball;
these states  are as follows:
$f_0^{bare}(1230^{+150}_{-\; 30}) $ and $f_0^{bare}(1600\pm
50) $.  To resolve this dilemma one needs
 the systematization of $q\bar
q$ states on the $(n,M^2)$ plot ($n$  being radial quantum number
of the meson and $M$  its mass): the systematization suggested
in \cite{Systematics} definitely proves $f_0^{bare}(1600\pm 50) $
to be  superfluous state on the $q\bar q$ trajectory. Correspondingly,
$f_0^{bare}(1230^{+150}_{-\; 30})$ and $ f_0^{bare}(1810\pm 30)$
should be  the $q\bar q$ states.

Below we present arguments
based on  the detailed
consideration of the $(n,M^2)$ plot,
while now let us discuss the variant which satisfies the constraints
given by the $q\bar q$ trajectories.

The nonet $2^3P_0q\bar q$ looks as follows:
\be
\begin{array}{cl}
2^3P_0q\bar q:&
f_0^{bare}(1230^{+150}_{-\; 30}),\; f_0^{bare}(1810\pm 30),\;
a_0^{bare}(1650\pm 50),\;K_0^{bare}(1885^{+\; 50}_{-100}),
\\
& \phi[f_0^{bare}(1230)]=40^{\circ}\pm 8^\circ\ .
\label{radnonet}
\end{array}
\ee
The $K$-matrix analysis \cite{YF}, together with previous
ones \cite{km,km1900}, enable to reveal in the
scalar-isoscalar sector the bare state     $f_0^{bare}(1600)$
 which is an extra
one for the nonet classification  $1^3P_0q\bar q$ and
$2^3P_0q\bar q$.
At the same time the couplings of $f_0^{bare}(1600)$
to the decay channels
$\pi\pi,K\bar K, \eta\eta, \eta\eta'$ obey the requirements imposed on
the glueball decay. This gives the reason to consider this state as the
lightest scalar glueball,
\be
0^{++}\; glueball: \qquad  f_0^{bare}(1600\pm 50)\ .
\ee
The lattice calculations are in an reasonable agreement with
such value of the lightest glueball mass \cite{lattice}.

After the onset of the decay channels the bare states
are transformed into
real resonances. For the scalar--isoscalar sector
we observe the following transitions by switching on the decay
channels:
\be
\begin{array}{cl}
f_0^{bare}(720)\pm 100) &\to \; f_0(980)\ ,\\
f_0^{bare}(1230^{+150}_{-\; 30}) &\to \;f_0(1300)\ , \\
f_0^{bare}(1260\pm 30) &\to \;f_0(1500)\ , \\
f_0^{bare}(1600\pm 50) &\to \;f_0(1420^{+150}_{-\;70})\ , \\
f_0^{bare}(1810\pm 30) &\to \;f_0(1750)\ .
\end{array}
\ee
The evolution of
bare states into real resonances is illustrated
by Fig. 2: the shifts of  amplitude poles in the complex-$M$
plane correspond to  gradual onset of the decay channels. Technically
it is done by  replacing  the phase spaces $\rho_a$ for
$a=\pi\pi,\pi\pi\pi\pi,K\bar K, \eta\eta, \eta\eta'$ in the
$K$-matrix amplitude
as follows: $\rho_a\to \xi \rho_a $, where the parameter
$\xi$ runs in the interval $0\le \xi \le 1$. At $\xi \to 0$ one has
bare states while the limit $\xi \to 1$ gives us the positions of
real resonances.

\subsection{ Overlapping of  $f_0$-resonances in the mass region
1200--1700 MeV: accumulation of widths of the $q\bar q$ states by the
glueball }

The occurrence of the broad resonance is not at all an accidental
phenomenon. It originated due to a mixing of states in the
 decay processes, namely, transitions
$f_0(m_1)\to real\; mesons\;
\to f_0(m_2)$. These transitions
result in  a specific phenomenon, that is, when several resonances
overlap one of them accumulates the widths of neighbouring resonances
and transforms into the broad state.

This phenomenon had been observed in
\cite{km,km1900} for scalar-isoscalar states,
and  the following scheme has been suggested in
\cite{glueball,ZPhys}: the broad state $f_0(1420^{+150}_{-\;70})$
is the descendant of the pure glueball which being in the neighbourhood
of $q\bar q$ states accumulated their widths and transformed into the
mixture of gluonium and $q\bar q$ states. In \cite{ZPhys}
this idea had been modelled for four resonances
$f_0(1300)$, $f_0(1500)$, $f_0(1420^{+150}_{-\;70})$ and $f_0(1750)$,
by using the language of the quark-antiquark and two-gluon states,
$q\bar q$ and $gg$: the decay processes were considered to be
transitions $f_0 \to q\bar q, gg$, correspondingly, the
same processes realized the mixing of the resonances.
In this model, the gluonium component was dispersed
mainly over three resonances,  $f_0(1300)$, $f_0(1500)$,
$f_0(1420^{+150}_{-\;70})$, so every  state is a mixture
of $q\bar q$ and $gg$ components, with roughly equal percentage
of gluonium (about 30-40\%).

 Accumulation of widths of overlapping resonances by one of them is a
well-known effect
in nuclear physics \cite{Shapiro,Okun,Stodolsky}.
In meson physics this phenomenon can play rather important role, in
particular, for exotic states which are beyond the $q\bar q$
systematics. Indeed, being among  $q\bar q$ resonances, the exotic
state creates a group of overlapping resonances.
The exotic state, which is not orthogonal
to its neighbours, after accumulating the "excess" of widths,
turns into the broad one. This broad resonance should be accompanied
by narrow states which are the descendants of states from which the
widths have been taken off. In this way, the existence of a broad
resonance accompanied by narrow ones may be a signature of the
exotics.  This possibility, in context of searching for exotic states,
was discussed in \cite{exotic,radius_YF}.

The broad state may be one of the components which forms the
confinement barrier:  the broad states after accumulating the widths of
neighbouring resonances play for these latter the role of locking
states. Evaluation of the mean radii squared of the broad state
$f_0(1420^{+150}_{-\;70})$ and its neighbouring resonances argues in
favour of this idea, for the radius of $f_0(1420^{+150}_{-\;70})$
is significantly larger than that for $f_0(980)$ and $f_0(1300)$
\cite{radius_YF, radius_PL} thus making it possible for
$f_0(1420^{+150}_{-\;70})$ to be locking state.

\subsection{Systematics of the scalar-isoscalar $q\bar q$ states on the
$(n,M^2)$ plot}

As is stressed above,
the systematics of $q\bar q$ states on the $(n,M^2)$ plot
argues that the broad state $f_0(1420^{+150}_{-70})$
and its predecessor, $f_0^{bare}(1600\pm 50)$, are  the states
beyond
$q\bar q$ classification.
Following \cite{Systematics}, we plot in Fig. 3a
the $(n,M^2)$-trajectories for $f_0$,
$a_0$ and $K_0$ states (the doubling of
$f_0$ trajectories is due to two flavour components, $n\bar n$ and
$s\bar s$). All trajectories are roughly linear,
and they clearly represent the
states with dominant $q\bar q$ component.  It is seen that one of the
states, either $f_0(1420^{+150}_{-70})$ or $f_0(1500)$, is superfluous
for $q\bar q$ systematics. Looking at the $(n , M^2)$-trajectories
for bare states (Fig. 3b), one can see that just $f_0^{bare}(1600)$ does
not fall on any linear $q\bar q$ trajectory. So it would be
natural to conclude that $f_0^{bare}(1600)$ is an exotic
state, i.e. the glueball.

Relying on the arguments given by
the systematics of $q\bar q$ states on the $(n,M^2)$ plot,
we can state that classification represented by (8) and (9) as a
basis for further analysis.
Lattice calculations support the solution (8)--(9):  calculations
give values for the mass of the lightest glueball in the interval
1550--1750 MeV \cite{lattice}.

\section{Hadronic decays, rules of quark combinatorics for couplings
and estimation of the quark-gluonium content of
resonances}

In this Section, on the basis of the quark combinatorics for the decay
coupling constants, we analyse
 quark-gluonium content of resonances $f_0(980),
f_0(1300), f_0(1500), f_0(1750)$ and $f_0(1420^{+150}_{-70})$. We
would like to bring attention of the reader to ambiguities which are
inherent in analyses  studying  hadronic decays of resonances only.

\subsection{Quark combinatorial relations for the decay couplings}

Within  $1/N$ leading-order terms \cite{t'Hooft},  hadronic decays
of meson resonances are determined by planar diagrams. An
example of the process is shown in Fig. 4a: the quarks flying
away from the initial $q\bar q$ state produce in a soft way (i.e. at
relatively large distances, $r\sim R_{confinement}$) the new pair of
light quarks ($u\bar u$, $d\bar d$, or $s\bar s$) and turn into white
hadrons, thus making it possible for initial quarks to leave the
confinement domain. In the limit of flavour SU(3) symmetry, the
production of all quarks is equivalent; still, heavier weight of
strange quark results in a suppression of the production
probability of $s\bar s $ pair. So the
following ratio of production probabilities takes place:
\be
u\bar u:d\bar d:s\bar s=1:1:\lambda\ ,
\ee
with $\lambda=$0.4---0.8 for meson decays
\cite{combinatorics,Klempt}. For
similar decays, the coupling constants (see Fig. 4) differ by the
coefficient which depends on the mixing angle of the $n\bar n$ and
$s\bar s$ components of the initial meson and parameter $\lambda$
\cite{combinatorics}. These coefficients are shown in Table 1 for
the decays $f_0 \to \pi\pi, K\bar K, \eta\eta, \eta\eta', \eta'\eta'$.

\begin{table}
\caption\protect{Coupling constants given by quark combinatorics for
$f_0$-meson and glueball
decaying into two pseudoscalar mesons in the leading terms
of  $1/N$ expansion;
$\varphi$ is the mixing angle for $n\bar n$ and $s\bar s$ states,
$f_0=n\bar n \cos\varphi+s\bar s \sin\varphi$,
and
$\Theta$ is the mixing angle for $\eta -\eta'$ mesons:
$\eta=n\bar n \cos\Theta-s\bar s \sin\Theta$ and
$\eta'=n\bar n \sin\Theta+s\bar s \cos\Theta$, where
$\cos\Theta\simeq 0.8$ and $\sin\Theta\simeq 0.6$.
\label{table1}}
\begin{tabular}{|c|c|c|c|}
~      &     ~                    &  ~                     &~           \\
~      & The $q\bar q$-meson decay&The glueball decay      &Identity  \\
~      & couplings in the         &couplings               & factor  \\
Channel& leading terms of $1/N$   &in the leading terms    &in phase \\
~      & expansion                &of $1/N$ expansion      &space    \\
~      &     ~                    &  ~                     &~ \\
\hline
~     &                          ~ & ~                      & ~    \\
$\pi^0\pi^0$
&$g\;\cos\varphi\; /\sqrt{2}$&$G\; /\sqrt{2+\lambda}$ &1/2 \\
 ~ & ~ & ~ & ~ \\
$\pi^+\pi^-$&$g\;\cos\varphi\; /\sqrt{2}$&$G\; /\sqrt{2+\lambda}$ &1 \\
~ & ~ & ~ & ~ \\
$K^+K^-$    &$g (\sqrt 2\sin\varphi+\sqrt\lambda\cos\varphi)/\sqrt 8 $ &
   $ G\; \sqrt{\lambda/(2+\lambda)} $                  &1 \\
~ & ~ & ~ & ~ \\
$K^0\bar K^0$ & $g (\sqrt 2\sin\varphi+\sqrt\lambda\cos\varphi)/\sqrt 8
$ & $G\; \sqrt{\lambda/(2+\lambda)}  $                            & 1
\\ ~ & ~ & ~ & ~ \\ $\eta\eta$ & $g\left
(\cos^2\Theta\;\cos\varphi/\sqrt 2+\right .$\hfill &$G\;
(\cos^2\Theta+\lambda\sin^2\Theta) \; /\sqrt{2+\lambda}$ & 1/2 \\ ~
&\hfill$\left .  \sqrt{\lambda}\;\sin\varphi\;\sin^2\Theta\right )$ &~
&\\ ~ & ~ & ~ & ~ \\ $\eta\eta'$ &
$g\sin\Theta\;\cos\Theta\left(\cos\varphi\; /\sqrt 2-\right .$\hfill
&$G\; \cos\Theta\; \sin\Theta\; (1-\lambda )\; /\sqrt{2+\lambda}$ & 1\\
~& \hfill $\left .\sqrt{\lambda}\;\sin\varphi\right ) $ &
\hfill &\\
~ & ~ & ~ & ~ \\
$\eta'\eta'$ &
$g\left(\sin^2\Theta\;\cos\varphi/\sqrt 2+\right .$\hfill
&$G\; (\sin^2 \Theta+\lambda\cos^2\Theta )\; /\sqrt{2+\lambda}$ &
1/2 \\
~&\hfill $\left .\sqrt{\lambda}\;\sin\varphi\;\cos^2\Theta\right)$ & ~
&\\ ~ & ~ & ~ & ~ \\ \end{tabular} \end{table}

The planar diagram for the transition  $glueball \to two\, mesons$ is
shown in Fig. 4b. Below, to illustrate the estimations, we consider
gluonium as the two-gluon composite system: the large value of the
soft-gluon mass, $m_{gluon}\sim 700-1000$ MeV \cite{gluon}, supports the
 model, though, we should stress, the results are of more
general meaning.

The planar diagram of  Fig. 4b represents
a two-stage transition $gluonium \;\to
q\bar q \to two\;  mesons$, and the second stage
is similar to the decay of $q\bar q$ meson.
Because of that the relations between couplings in the transition
$gluonium \to two\; pseudoscalars$ are governed by the magnitudes given
in Table 1, with fixed values of mixing angle for the $n\bar n$ and
$s\bar s$ components, $q\bar q=n\bar n\cos \varphi_G +s\bar s\sin
\varphi_G $, which have been formed in the process $gluonium \to q\bar
q$. The angle $\varphi_G$ is determined by the parameter $\lambda_G$
entering  the first stage of the process $gluonium\to q\bar q$ with
 relative probability $u\bar u:d\bar d:s\bar s=1:1:\lambda _G$.
Then
\be
\cos \varphi_G=\sqrt{\frac2{2+\lambda_G}}\ .
\ee
The
relations between coupling constants for the decays $gluonium \to
\pi\pi, K\bar K, \eta\eta, \eta\eta', \eta'\eta'$ are given in Table 1
as well. In principle, $\lambda_G$ may differ from the suppression
parameter inherent to transitions $q\bar q\to  two\; mesons$. However,
it looks reasonable to use, as the first approximation, the same value
of $\lambda$ for both stages of the gluonium decay, because the
coefficients for the transitions  $q\bar q\to  two\; mesons$ change
rather weakly with $\lambda$ belonging to the interval 0.4--0.8.

The sums of couplings squared for the decay transitions of
the $q\bar q$ meson and gluonium are of the same order, as it
follows from the rules of the $1/N$-expansion
\cite{t'Hooft} ($N=N_c=N_f$ where $N_c$ and $N_f$ are numbers of
colours and flavours). Let us denote the sum of couplings squared
for transitions of Fig. 4a-type, $q\bar q-state \to \sum mesons$,
as $g_{q\bar q}^2$ and the corresponding value for the gluonium
decay, $gluonium \to \sum mesons$, as $g^2_{gluonium}$. The values
$g_{q\bar q}^2$ and $g^2_{gluonium}$ can be represented as
discontinuities of the self-energy diagrams Figs. 4c,d, with
cuttings shown by dashed lines: the cut block stand for
the couplings shown in Figs. 4a,b. In  terms of the
$1/N$-expansion, the diagram Fig.  4c (and $g_{q\bar q}^2$) is of the
 order of
\be
g_{q\bar q}^2 \sim G^2_{q\bar q-meson\to q\bar q}N_f \sim
\frac{N_f}{N_c}\ ,
\ee
because $G^2_{q\bar q-meson\to q\bar q} \sim 1/N_c$.
Likewise, for $g^2_{gluonium}$ determined
by diagrams of  Fig. 4d-type, one has:
\be
g_{gluonium}^2 \sim G^2_{gluonium\to gg}N^2_f \sim
\frac{N_f^2}{N_c^2}\; .
\ee
The coupling for the transition
$gluonium \to two-gluon\, state$, $G_{gluonium\to gg}$, is
of the order of $G_{gluonium\to gg} \sim 1/N_c$.
The estimates of  couplings $G_{q\bar q-meson\to q\bar q}$ and
$G_{gluonium\to gg}$ are done by using  basic self-energy
diagrams for composite systems, which are  of the order of
unity: in cases under consideration such are the diagrams of Figs.
4e,f.

The two-stage nature of the decay $gluonium \to two\; pseudoscalars$
is a source of ambiguities in the determination of the quark-gluon
content of mesons, if we restrain ourselves by hadronic decays only.
The matter is that the $q\bar q$ meson with the quark content $n\bar
n\cos \varphi +s\bar s\sin \varphi $ at $\varphi\simeq\varphi_G$ has
the same correlations between coupling constants for the transitions
 $f_0 \to \pi\pi, K\bar K, \eta\eta, \eta\eta'$ as those for the
glueball. At $0.4 \le\lambda_G\la 0.8$ we have $24^\circ \la
\varphi_G\la 32^\circ$: it means that the analysis of hadronic decays
cannot distinguish between $q\bar q$ state with $\varphi \simeq
\varphi_G$ and  true gluonium.

\subsection{Decay couplings for the resonances
 $f_0(980)$, $ f_0(1300)$, $f_0(1500)$, $f_0(1750)$,
$f_0(1420^{+150}_{-70})$ into channels $\pi\pi, K\bar
K,\eta\eta,\eta\eta', \pi\pi\pi\pi$ }

The $K$-matrix fit to the data gives us directly the characteristics of
the bare states only. To extract resonance parameters one
needs additional calculations to be carried out with the obtained
amplitude. The couplings for
resonance decay are
extracted by calculating residues of the
amplitude poles related
 to the resonances \cite{width}. In a more detail, the
amplitude $A_{a\to b}$, where $a$,  $b$ mark the
channels $\pi\pi, K\bar K,\eta\eta,\eta\eta', \pi\pi\pi\pi$,
can be written near the pole as:
\be
A_{ab}\simeq \frac {g_a^{(n)}g_b^{(n)} }{\mu_n^2-s}e^{i(\theta_a^{(n)}
+\theta_b^{(n)})}+B_{ab}\ .
\label{pole}
\ee
The first term in (\ref{pole}) represents the pole singularity
and the second one, $B_{ab}$, is a smooth background.
The pole position $s=\mu_n^2$
determines the mass of the resonance, with total width  $\mu_n=
M_n-i\Gamma_n /2$,
and the real factors $g_a^{(n)}$ and $g_b^{(n)}$ are the decay coupling
constants of the resonance to channels $a$ and $b$.
The couplings $g_a^{(n)}$ given in Table 2
stand for the solution
with glueball in the vicinity of 1600 MeV; also are shown
 the couplings for the predecessor bare states.

The couplings of Table 2 demonstrate a strong change of couplings
during the  evolution from bare states to real resonances. Note that the
change occurs not only in  absolute values of couplings but
in relative magnitudes as well. The resonances $f_0(1300)$, $f_0(1500)$,
and $f_0(1750)$  demonstrate a reduction of  relative weight of
the coupling constant squared, $g_{K\bar K}^2$, while the same coupling
in the broad state $f_0(1420^{+150}_{-70})$ increases.

\begin{table}
\caption\protect{ Couplings squared, $g^2_{a}$,
(in GeV$^2$ units) for bare states
and their resonance-descendants
\label{table2}   }
\begin{tabular}{|c|ccccc|c|}
\hline
~&~&~&~&~&~&~\\
State  &$g^2_{\pi\pi}$&$g^2_{K\bar K}$&$g^2_{\eta\eta}$
&$g^2_{\eta\eta'}$&$g^2_{\pi\pi\pi\pi}$&$\sum g^2_a$\\
\hline
~&~&~&~&~&~&~\\
$f_0^{bare}(650^{+12}_{-30})$ &0.167&0.528 &0.06  &- &0    &0.755 \\
$f_0(980)                  $ &0.076 &0.186 &0.072 &- &0.004&0.338 \\
\hline
~&~&~&~&~&~&~\\
$f_0^{bare}(1220^{+15}_{-30})$ &0.083 &0.099 &0.025 &- &0.517&0.724 \\
$f_0(1300)                 $ &0.026 &0.002 &0.003 &- &0.132&0.163 \\
\hline
~&~&~&~&~&~&~\\
$f_0^{bare}(1265^{+15}_{-45})$ &0.553&0.271&0.045 &0.032&0    &0.881\\
$f_0(1500)                 $ &0.038 &0.009 &0.007 &0.006&0.074&0.134 \\
\hline
~&~&~&~&~&~&~\\
$f_0^{bare}(1820\pm 40)$ &0.059&0.019&0.001 &0.043&0.262&0.384\\
$f_0(1750)                 $ &0.086 &0.003 &0.009 &0.028&0.117&0.243 \\
\hline
~&~&~&~&~&~&~\\
$f_0^{bare}(1585^{+10}_{-45})$ &0.124&0.062&0.025 &0.008&0.924&0.513\\
$f_0(1420^{+150}_{-70})$ &0.304 &0.271 &0.062 &0.016&0.382&1.035 \\
\hline
\end{tabular}
\end{table}

The growth of relative weight of  $g_{K\bar K}^2$
in $f_0(1420^{+150}_{-70})$, that is the glueball descendant, can be
unambigously interpreted: by accumulating the widths of neighbouring
resonances this one acquires a noticeable $q\bar q$ component,
with a large amount of $s\bar s $ state.

Let us look at what the quark combinatorics rules given in Table
1 tell us about the proportion of $s\bar s$, $n\bar n$ and gluonium
components in the studied resonances. The
coupling constants squared for $f_0 \to \pi\pi, K\bar K, \eta\eta,
\eta\eta'$ can be written as follows:
\be
g^2_{ \pi\pi}=\frac 32\left (\frac{ g}{\sqrt 2}\cos\varphi
+\frac{ G}{\sqrt{2+\lambda}} \right )^2,
\label{gpipi}
\ee
$$
g^2_{K\bar K} =2
\left (\frac{g}{2} (\sin\varphi+\sqrt{\frac{ \lambda}{2}}\cos\varphi)
+G \sqrt{\frac{\lambda}{2+\lambda}}\; \right )^2
$$
$$
g^2_{\eta\eta}=\frac 12 \left (g\; (\frac{\cos^2\Theta}{\sqrt
2}\cos\varphi +\sqrt{\lambda}\;\sin\varphi\;\sin^2\Theta\; )+
\frac{G}{ \sqrt{2+\lambda}}\;
 (\cos^2\Theta+ \lambda\sin^2\Theta\ )\right )^2
$$
$$
g^2_{\eta\eta'} =\sin^2\Theta\;\cos^2\Theta\; \left (
g\;  (\frac{1}{\sqrt 2}\cos\varphi-
\sqrt{\lambda}\;\sin\varphi  ) +G\;
\frac{1-\lambda}{\sqrt{2+\lambda}}
 \right )^2\ .$$

The term proportional to $g$ is responsible for the transition $q\bar
q\to two\; mesons$, while that proportional to $G$ stands for the
transition $gluonium\to two\; mesons$. Correspondingly, the magnitudes
$g^2$ and $G^2$ are proportional to probabilities to find the $q\bar q$
and gluonium components in the considered meson.

First of all, let us determine the mean value of mixing angle
for $n\bar n$/$s\bar s$ components in the intermediate state,
$<\varphi>$:
\be
f_0\to gluonium+q\bar q \to n\bar n \cos <\varphi> +s\bar s \sin
<\varphi> \to two\; mesons.
\ee
We define $<\varphi >$ as the angle for the
coupling constants squared (\ref{gpipi}) at $G=0$.  Then we have
for the studied resonances:
\bea
&f_0(980):\quad &<\varphi >\simeq
-67^\circ\ , \\ \nonumber
&f_0(1300):\quad &<\varphi >\simeq -5^\circ\ ,
\\ \nonumber
&f_0(1500):\quad &<\varphi >\simeq 8^\circ\ ,
\\ \nonumber
&f_0(1420^{+150}_{-70}):\quad &<\varphi >\simeq 37^\circ\ ,
\\ \nonumber
&f_0(1750):\quad &<\varphi >\simeq -27^\circ\ .
\eea
The value $<\varphi [f_0(1420^{+150}_{-70})]>\simeq 37^\circ$ is
very close to the mixing angle of the flavour singlet state,
$\varphi_{singlet}= 35.3^\circ$. We see that $f_0(1420^{+150}_{-70})$
is a mixture of the gluonium and $(q\bar q)_{singlet}$.

Generally, by
fitting formulae (\ref{gpipi}) to the coupling constants squared
of Table 2,
one can determine the mixing angle $\varphi$ as a
function of $G/g$.
The curves in Fig. 5 demonstrate the dynamics of
$\varphi$ with respect to the ratio $G/g$ for the resonances
$f_0(980)$, $f_0(1300)$, $f_0(1500)$, and $f_0(1750)$.

The magnitudes $g^2$ and $G^2$ are proportional, correspondingly, to the
probabilities for quark/gluonium components, $W_{q\bar q}$ and
$W_{gluonium}$, to be in the considered resonance:
\be
g^2=g^2_{q\bar q} W_{q\bar q}\; , \qquad
G^2=g^2_{gluonium}W_{gluonium}\ .
\ee
According to the
rules of $1/N$ expansion \cite{t'Hooft}, the coupling constants,
$g^2_{q\bar q}$  and    $g^2_{gluonium}$,  are of the
same order (see Section 3.1), therefore  we accept as a rough
estimation:
\be
G^2/g^2=W_{gluonium}/W_{q\bar q}\ .
\label{G/g}
\ee
In Fig. 5 we vary
$G/g$ in the interval $-0.8 \le G/g \le0.8$; that,
in accordance with (\ref{G/g}),    is related,
  to a possible admixture of the gluonium
component up to 40\%: $W_{gluonium}\le 0.40$.

The $q\bar q$ components in the
resonances $f_0(1300)$, $f_0(1500)$ reveal rather moderate change
of $\varphi$ versus a persentage of gluonium component:
\be
W_{gluonium}[f_0(1300)]\le 40\% \; : \qquad
-25^\circ \le \varphi [f_0(1300)] \le 13^\circ\ ,
\label{phi1300}
\ee
and
\be
W_{gluonium}[f_0(1500)]\le 40\%\; : \qquad
-3^\circ \le \varphi [f_0(1500)] \le 17^\circ\ .
\label{phi1500}
\ee
More sensitive to the glueball admixture is
the $q\bar q$ components in  $f_0(1750)$:
\be
W_{gluonium}[f_0(1750)]\le 40\%\; : \qquad
-55^\circ \le \varphi [f_0(1750)] \le -2^\circ \; .
\label{phi1750}
\ee
The $n\bar n/s\bar s$ ratio
in $f_0(980)$ is also rather sensitive to the presence of
gluonium component. However, for this
case it is hardly possible to assume the glueball admixture to be more
than 20\% \cite{phi}. Correspondingly, we have
\be
W_{gluonium}[f_0(980)]\le 20\%\; : \qquad
-95^\circ \le \varphi [f_0(980)] \le -40^\circ.
\label{phi980}
\ee
The performed analysis clearly demonstrates the impossibility to fix
unambigously the $n\bar n$, $s\bar s$ and gluonium components in the
resonances
$f_0(980)$, $f_0(1300)$, $f_0(1500)$,  $f_0(1750)$
by using their couplings  to hadronic
channels only. For more understanding of the structure of these  mesons,
one needs additional data.

Still, the $K$-matrix analysis
\cite{YF} provides us with a bulk of information
on meson states; that   makes it possible to trace the evolution
of  coupling constants from bare
state, $f_0^{bare}$, to real resonances.

\section{Evolution of coupling constants for $f_0 \to
\pi\pi$, $K\bar K$, $\eta\eta$, $\eta\eta'$ at the onset of the decay
channels}

In this section, basing on the $K$-matrix analysis results
\cite{YF}, we study
the dynamics of parameters for  the resonances
$f_0(980)$, $f_0(1300)$, $f_0(1500)$,  $f_0(1750)$,
$f_0(1420^{+150}_{-70}) $ by switching on/off
 gradually the decay channels.

\subsection{The onset of decay channels in the $K$-matrix amplitude}

When the decay channels are switched off, the coupling constants to
two-meson channels are determined by the residues of the $K$-matrix
poles, while, after switching them on, the coupling constants are
determined by the residues of amplitude poles. Let us clarify this point
 in more detail.

Scattering amplitude of the two pseudoscalar mesons had been fitted in
\cite{YF} in the form:
\be
\hat A= \hat K \frac{I}{I-i\hat \rho\hat K}\ ,
\label{4.1}
\ee
where $\hat K$ is the $5\times 5$-matrix for five channels under
investigation ($\pi\pi, K\bar K,\eta\eta,\eta\eta', \pi\pi\pi\pi$). The
$\hat K$-matrix is real and symmetrical $K_{ab}(s)=K_{ba}(s)$, $I$ is
the unit matrix, $I= diag(1,1,1,1)$, and $\hat \rho $ is the
diagonal matrix for phase space,
$\hat \rho=diag (\rho_{\pi\pi}$,$\rho_{ K\bar
K}$,$\rho_{\eta\eta}$,$\rho_{\eta\eta'}$,
$\rho_{\pi\pi\pi\pi})$.

The $K$-matrix element was represented in \cite{YF} as a sum of pole
terms and smooth background $f_{ab}(s)$:
\be
K_{ab}(s)=\sum_n\frac {g_a^{(n)}g_b^{(n)} }{\mu_n^2-s}+f_{ab}(s)\ ,
\label{4.2}
\ee
$\mu_n$ being the mass of  bare state and $g_a^{(n)}$  the coupling
 of $f_0^{(bare)}(\mu_n)$ to the channel $a$.

Fitting to data performed in \cite{YF} fixes the
parameters of the $K$-matrix amplitude (coupling constants
$g_{ab}(s)$, masses of the $K$-matrix poles $\mu_n$ and regular terms
$f_{ab}$). With the completely determined parameters we can investigate
the dynamics of the onset of decaying processes.

Such an investigation of the resonance evolution
was suggested in \cite{AS-GAMS}, where rather simple variant
had been considered: in the amplitude (\ref{4.1}) the substitution
$\hat \rho \to x\rho$ has been done, with $x$ varying in the interval
$0 \le x \le 1$. Thus constructed  amplitude $\hat A(x)$
gives us the real amplitude at $x=1$,  and at
$x\to 0$ the amplitude $\hat A(x)$ turns into the K-matrix one,
$\hat A(x\to 0)\to \hat K$, which is the amplitude for  bare states.
Generally, the onset of the decay channels in the $K$-matrix
amplitude can be investigated by substituting parameters as follows:
\be
g_a^{(n)} \Rightarrow  \xi_n(x) g_\alpha^{(n)}\ , \qquad
f_{ab} \Rightarrow  \xi_f(x) f_{ab}\ .
\label{4.3}
\ee
Here the parameter-functions $\xi_n(x)$, $\xi_f(x)$ obey the
requirements $\xi_n(0)=\xi_f(0)=0$ and $\xi_n(1)=\xi_f(1)=1$.
A simple variant studied in \cite{AS-GAMS} corresponds to
$\xi_n(x)=\sqrt{x}$ and $\xi_f(x)=x$.

The parameters $\xi_n(x)$ and $\xi_f(x)$ control the dynamics of
the onset of the decay channels for resonances. This dynamics may be
different for  different states, say, $q\bar q$ state and  gluonium.
Below we use $\xi_n(x)=\sqrt{x}$ for the states connected with
$f_0(980)$, $f_0(1300)$, $f_0(1500)$,  $f_0(1750)$, while for the broad
state the dependence $\xi_n(x)=x^{1/4}$ is accepted. For the background
term we use $\xi_n(x)=x$.

\subsection{Evolution of the decay couplings}

Below the evolution of states related to the resonances
$f_0(980)$, $f_0(1300)$, $f_0(1500)$,  $f_0(1750)$, and
$f_0(1420^{+150}_{-70}) $ will be considered one by one. The
procedure is as follows: the substitution (\ref{4.3}) is done and $x$ takes
the numbers
  $x=0.1,0.2,...,0.8,0.9$. The value $x=0$
corresponds to bare states, and the couplings and masses had been found
as fitting parameters in
\cite{YF}; for $x=1$ calculations were performed in
\cite{width} and the couplings are presented in Table 2.
Furthermore, for different but fixed $x$ we find the position
of pole and calculate the residues of the amplitudes thus determining
$g_a^{(n)}(x)$.

\subsubsection{Resonance $f_0(980)$}

For the $f_0(980)$
the normalized couplings $\gamma_a= g_a/\sqrt{g^2_{\pi\pi}+g^2_{K\bar
K}}$, where $a=\pi\pi, K\bar K$ are shown in Fig. 6a.
Correspondent poles at different $x$
are placed on the trajectory $f_0^{bare}(720)-f_0(980)$ (see Fig. 2).
Looking at Fig. 6a, one can see that at small $x$
$g^2_{\pi\pi}<g^2_{K\bar K}$, that is natural, for the state
$f_0^{bare}(720)$ is close to the flavour octet. With the increase of
$x$, the coupling constants become equal to each other,
and in the interval $0.6\le x \le 0.8$
the coupling to pion channel is
greater than  to kaon channel, $g^2_{\pi\pi}(x\sim
0.7)>g^2_{K\bar K}(x\sim 0.7) $, thus revealing a relative reduction of
$s\bar s$ component. However, at $x\sim 0.8$, when the amplitude pole
approaches the $K\bar K$ threshold, see Fig. 2,  the relative weight of the
$K\bar K$ channel is strenthened to certain extent.

For every fixed $x$, the formula (16) has been fitted to the values of
$g^2_{\pi\pi}(x)$
and $g^2_{K\bar K}(x)$ in order to find out $\varphi$ as a function
of $G/g$: a set of obtained curves is shown in Fig. 6b. At $x=0$ the curve
$\varphi(G/g)$ corresponds to maximal possible values of mixing angles;
with the increase of $x$ these curves move smoothly to smaller $|\varphi|$,
while at $x>0.8$ they turn backward rather sharply. The curve for
to $x=1$ is also shown in Fig. 5, it coincides with a good accuracy with
that at $x=0$.

In Fig. 6b there are also two curves for $\varphi$ with
gradual accumulation of the gluonium component which results in
$W_{gluonium}= 20\%$ at $x=1$: recall, we suggest
$W_{gluonium}\le 20\%$. These are dotted curves which originate from the
point $G/g=0$ at $x=0$
(in the K-matrix analysis \cite{YF} the bare state $f_0^{bare}(720)$
is the pure  $q\bar q$) but furthermore they drift to $G/g>0$ and
$G/g<0$ and end at $x=1$ with $G/g=0.45$ and $G/g=-0.45$,
correspondingly. These final points stand for $W_{gluonium}=20\%$.
Respectively, the variations of $\varphi$ are shown separately by Fig.
6c; the area inbetween the curves is just the region of reasonable
values of $\varphi$ with the evolution of the decay widths. The lower
curve tells us that the reduction of $g^2_{K\bar K}$ with the increase
of $x$ to the region $x<0.8$ is plausible due to the growth of the
gluonium component leaving the ratio $n\bar n/s\bar s$
approximately constant.
The upper curve related to
$G/g<0$ testifies the possible evolution when,
in parallel with the increase
of the gluonium component, the weight of the $s\bar s$ component
becomes smaller. But at $x>0.8$, both curves
demonstrate sharp increase of the $s\bar s$ component.

At $x=1$, when mixing angle $\varphi$ which defines the
quark content of the $f_0(980)$, $(n\bar n\cos\phi+s\bar s\sin\phi)$,
its value may vary  in the intervals $ -90^\circ \le \varphi \le
-48^\circ $ and $ 85^\circ \le \varphi \le 90^\circ $ (Fig. 6c). It
would be instructive to compare these values with mixing angles
obtained for radiative decays $\phi (1020)\to \gamma f_0(980)$ and
$f_0(980) \to \gamma\gamma$. The combined analysis \cite{phi} for these
decays provided us with two possible solutions:  $\varphi= -48^\circ
\pm 6^\circ $ and $\varphi= 86^\circ \pm 3^\circ $.

The areas for $\varphi[f_0(980)]$ allowed by  radiative and hadronic
decays are shown in Fig. 7. One can see that the constraints  for
mixing angle obtained from the study of hadronic decays of the
$f_0(980)$ are in a nice agreement with the values obtained from the
study of radiative decays.  However, one should  stress, that we cannot
expect from  hadronic processes more rigid limitations for mixing
angle of the $f_0(980)$ than those known from  radiative decays.

\subsubsection{Resonances  $f_0(1300)$, $f_0(1500)$ and  $f_0(1750)$}

For the resonances $f_0(1300)$, $f_0(1500)$ and $f_0(1750)$,
the fraction of the $s\bar s$ component decreases.
They flow away from these resonances and enter the broad state
$f_0(1420^{+150}_{-70}) $.
Figure 8 demonstrates normalized coupling constants
squared $\gamma_a= g_a/\sqrt{\sum_b g^2_b}$ as a function of $x$. One
can see that for $f_0(1300)$, $f_0(1500)$ the values $\gamma_{K\bar K}$
are falling down while $\gamma_{\pi\pi}$ increase. At the
same time, for the broad state the normalized coupling $\gamma_{K\bar
K}$ is growing up.

The description of ratios $\gamma_a$ by Eq. (16) is shown
in Fig. 8: one can see that quark combinatorial rules describe
reasonably a number of data on the decay coupling constants.
But, as was stressed above, Eq.
(\ref{gpipi}) does not define the content of
a resonance, providing a correlation
only between mixing angle $\varphi$ and
ratio $G/g$. These correlations are presented in Figs. 9a,b,c
for the $f_0(1300)$, $f_0(1500)$,  $f_0(1750)$ at various $x$. Implying
 the predecessors of these resonances to be pure $q\bar q$ states,
we display the correlations $(\varphi,G/g)$
 at $G^2/g^2 \le 0.4$ that corresponds to (\ref{phi1300}),
(\ref{phi1500}) and (\ref{phi1750}).  Dotted curves in Figs. 9a,b,c
stand for $(\varphi,G/g)$ related to a maximal capture of the gluonium
component by these resonances.

\subsubsection{Broad state $f_0(1420^{+150}_{-70}) $ }

The evolution from the $f_0^{(bare)}(1600)$ to $f_0(1420^{+150}_{-70}) $
is accompanied by the accumulation of  $q\bar q$ component and
the growth of ratio $g/G$.
In this way, Fig. 9c demonstrates the
correlation $(\varphi,G/g)$: for small $x$,
when parameters of the corresponding state are close to those of
$f_0^{(bare)}(1600)$, the ratio $g/G$ is small. We see
that for the broad state the mixing angle $\varphi$ does not depend
practically on $g/G$ at $g^2/G^2 \le 0.50$.

The value of the mixing
angle at $x=1$, $\varphi =37^\circ$, proves that the
$f_0(1420^{+150}_{-70}) $ can accumulate the flavour singlet
component of $q\bar q$ only; that perfectly agrees with
its gluonium origin.

Figure 10 demonstrates the change of $\varphi$ with increasing $x$:
this curve does not depend on the rate of accumulation of the
$(q\bar q)_{singlet}$ component.

\section{Conclusion}

We have performed the analysis of coupling constants for the resonances
$f_0(980)$, $f_0(1300)$, $f_0(1500)$, $f_0(1750)$,
and the broad state $f_0(1420^{+150}_{-70}) $ to channels
$\pi\pi, K\bar K, \eta\eta, \eta\eta'$  as well as
observed the evolution of
bare states into these resonances by switching on/off the decay channels
(see (10) and Fig. 2). Our
analysis has been based on  Solution II-2 of
the paper \cite{YF}; in this solution, the bare state $f^{(bare)}_0(1600\pm
50)$ is the candidate for the glueball.

During the evolution of states, the coupling constants $f_0\to hadrons$
change considerably not only in magnitude but relative weight as well,
that is due to a strong mixing of states, because of the decay
processes $f_0(m_1) \to real\, mesons  \to f_0(m_2)$.  Using the language
of bare states , $f^{(bare)}_0$,  a  rigid classification can be
established for $q\bar q$ states; however, for real resonances the
presence of the gluonium component results in certain
uncertainties.

For the discussed resonances the results are as follows.

{\bf 1. $f_0(980)$}: This resonance is dominantly the $q\bar q$ state, the
admixture of the glueball component is not more than 20\%, $W_{gluonium}
\le 0.20$.  Rather large $s\bar s$ component
is also present. Taking
account of the representation $q\bar q=n\bar n \cos \varphi+s\bar s
\sin \varphi$, the hadronic decays give us the following constraints:
$-90^\circ \le \varphi \le -40^\circ$    or
$85^\circ \le \varphi \le 90^\circ$, which
are in agreement with  data on radiative
decays $f_0(980) \to \gamma\gamma$ and $\phi(1020) \to\gamma f_0(980)$
\cite{phi} (see Fig. 7). Rather large uncertainties in the determination of
mixing angle  are due to the sensitivity of coupling constants to
plausible
small admixtures of the gluonium component. When
the gluonium component is absent,
hadronic decays provide $\varphi =<\varphi>=-67^\circ$.

{\bf 2. $f_0(1300)$ and $f_0(1500)$}: These resonances are the
descendants of bare $q\bar q $ states
$f^{(bare)}_0(1230^{+150}_{-30})$ and $f^{(bare)}_0(1260 \pm 30)$ which
both of them are  flavour singlets. The resonances $f_0(1300)$ and
$f_0(1500)$ are formed due to a strong mixing with gluonium
state $f^{bare}_0(1600)$  as well as with one another.  The $q\bar q$
content, $q\bar q=n\bar n \cos \varphi+s\bar s\sin \varphi$, in both
resonances strongly depends on the admixture of the gluonium component.
At $W_{gluonium} \le 0.40$ the mixing angles change, depending on
$W_{gluonium}$, in the intervals
$-25^\circ \le \varphi[f_0(1300)] \le 13^\circ$    and
$-3^\circ \le \varphi[f_0(1500)] \le 17^\circ$.

{\bf 3. $f_0(1750)$}: This resonance is the descendant of bare state
$2^3P_1 q\bar q$,          $f^{(bare)}_0(1810 \pm 30)$, and this bare
state has the flavour wave function close to the octet one. During
evolution and mixing (presumably with the gluonium) the quark
component, $q\bar q=n\bar n \cos \varphi+s\bar s\sin \varphi$, can
change significantly:  with $W_{gluonium} \le 0.2$, we have $-45^\circ
\le \varphi[f_0(1750)] \le -10^\circ$. Therefore, $f_0(1750)$ keeps his
large $s\bar s $ component, and rather small coupling constant
$f_0(1750)\to K\bar K$ should not mislead
us, for the production of $K\bar K$
is suppressed at $\varphi \sim -30^\circ$, see Table 1 and Eq.
(\ref{gpipi}).

{\bf 4. $f_0(1420^{+150}_{-70})$}: This broad state is the descendant of
the $f^{(bare)}_0(1600 \pm 50)$ which we believe to be the glueball. The
analysis of hadronic decays  of this resonance confirms the glueball nature
of this resonance: the $q\bar q$ component is allowed to be in the flavour
singlet only, $ \varphi[f_0(1420^{+150}_{-70})] \simeq 37^\circ$, though the
value of the possible admixture of $(q\bar q)_{singlet}$ cannot be fixed by
 hadronic decays. In this way, the $f_0(1420^{+150}_{-70})$ is a mixture
$gluonium + (q\bar q)_{singlet}$; the impossibility to find out the
quark-antiquark component is due to the fact that correlations between
the decay coupling constants  are the same for the gluonium and $q\bar q$
singlet.

\section*{Acknowledgement} We are indebted to A.V. Anisovich, D.V. Bugg,
L.G. Dakhno for useful remarks concerning  related problems. The work is
supported by the RFFI grant N  01-02-17861.

\newpage
\begin{figure}
\centerline{\epsfig{file=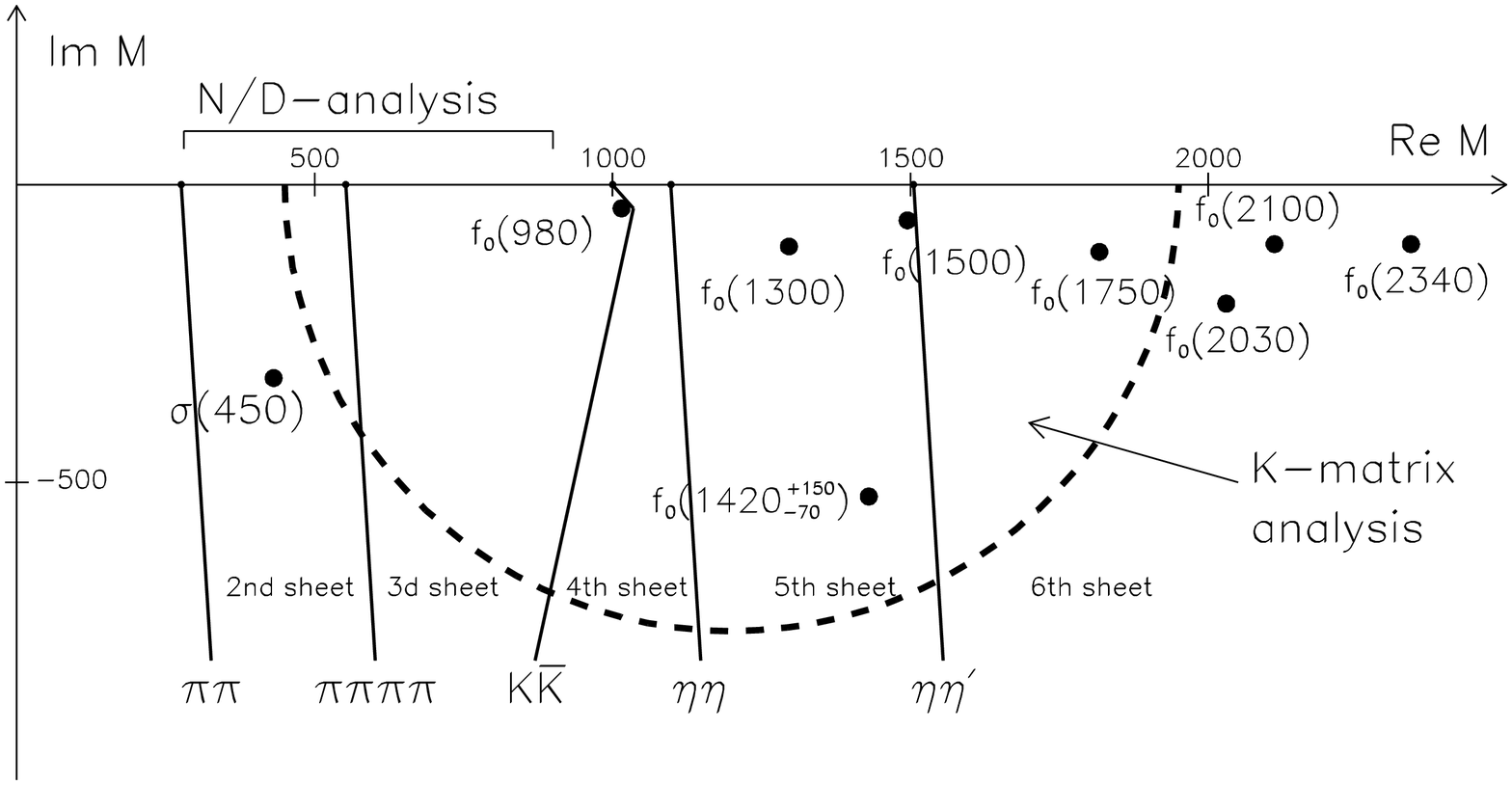,width=14cm}}
\caption\protect{ Complex $M$-plane for the $(IJ^{PC}=00^{++})$ mesons.
Dashed line encircles the part of the plane where the $K$-matrix analysis
[1] reconstructs analytical $K$-matrix amplitude: in this area the poles
corresponding to resonances $f_0(980)$, $f_0(1300)$, $f_0(1500)$,
$f_0(1750)$ and the broad state $f_0(1420\;^{+\;150}_{-70})$ are
located. Beyond this area, in the low-mass region, the pole
of the light $\sigma$-meson is located
(shown by the point the position of pole,
$M=(430-i320)$ MeV,
corresponds to the
result of $N/D$ analysis [11];
the crossed bars stand for $\sigma$-meson pole found in [19]). In the
high-mass region, one has resonances $f_0(2030),f_0(2100),f_0(2340)$
[17,18].  Solid lines statnd for the cuts related to the thresholds
$\pi\pi,\pi\pi\pi\pi,K\bar K,\eta\eta,\eta\eta'$.}
\end{figure}

\newpage
\begin{figure}
\centerline{\epsfig{file=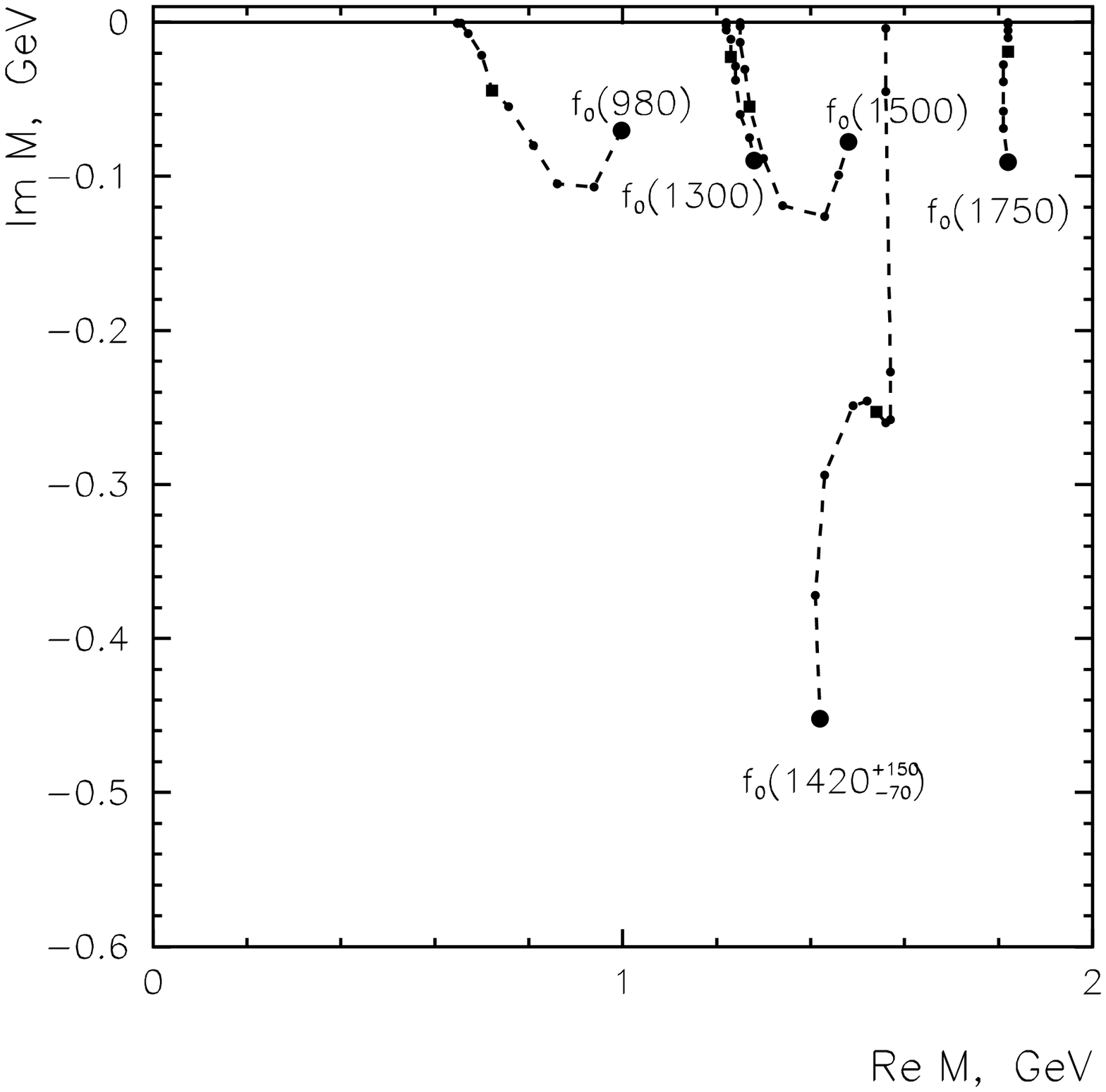,width=12cm}}
\caption\protect{ Complex $M$-plane: trajectories of the poles
for $f_0(980)$, $f_0(1300)$, $f_0(1500)$,
$f_0(1750)$, $f_0(1420\; ^{+\;150}_{-70})$
during gradual onset of the decay processes.}
\end{figure}

\newpage
\begin{figure}
\centerline{\epsfig{file=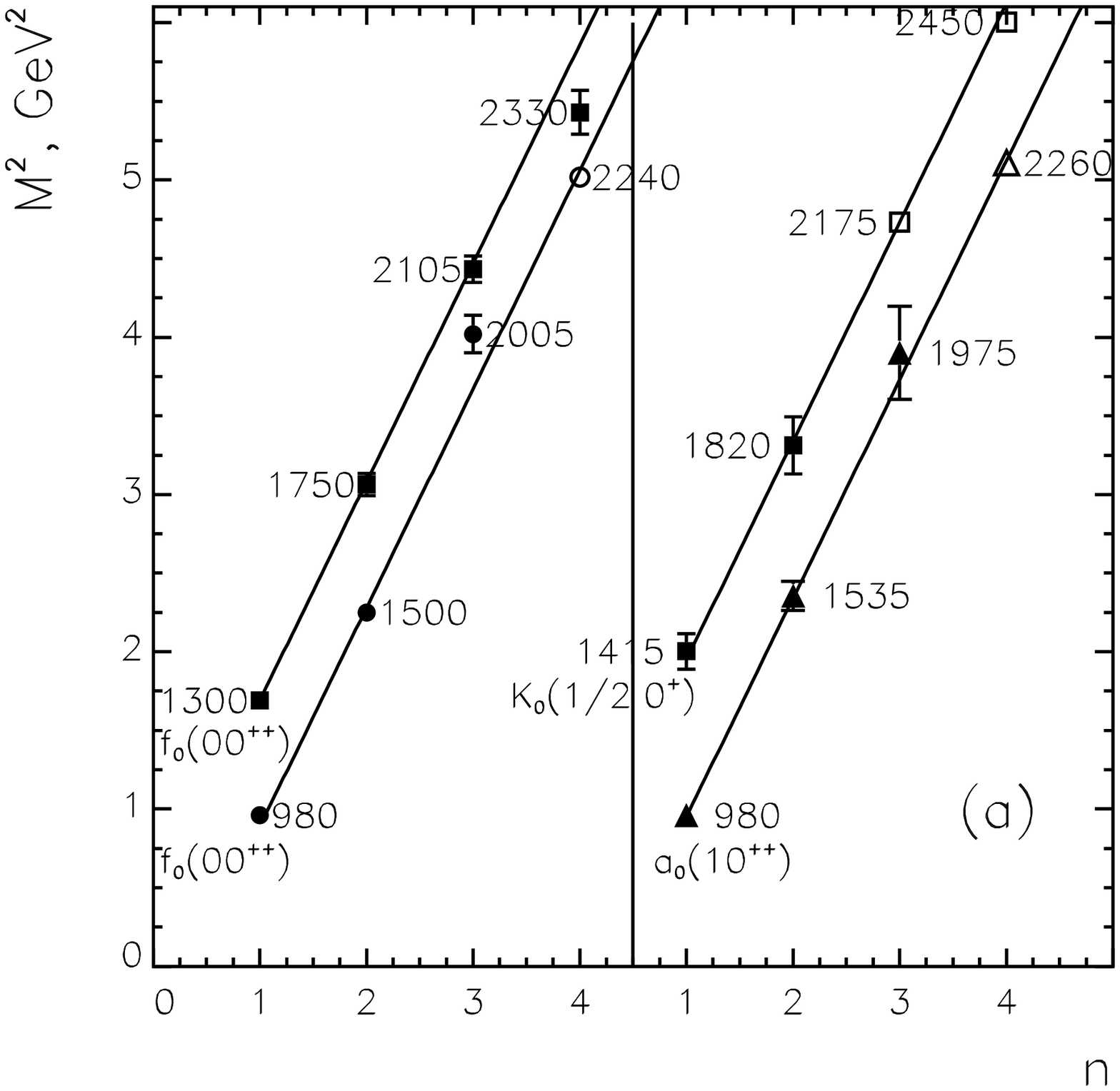,width=9cm}}
\vspace{-0.5cm}
\centerline{\epsfig{file=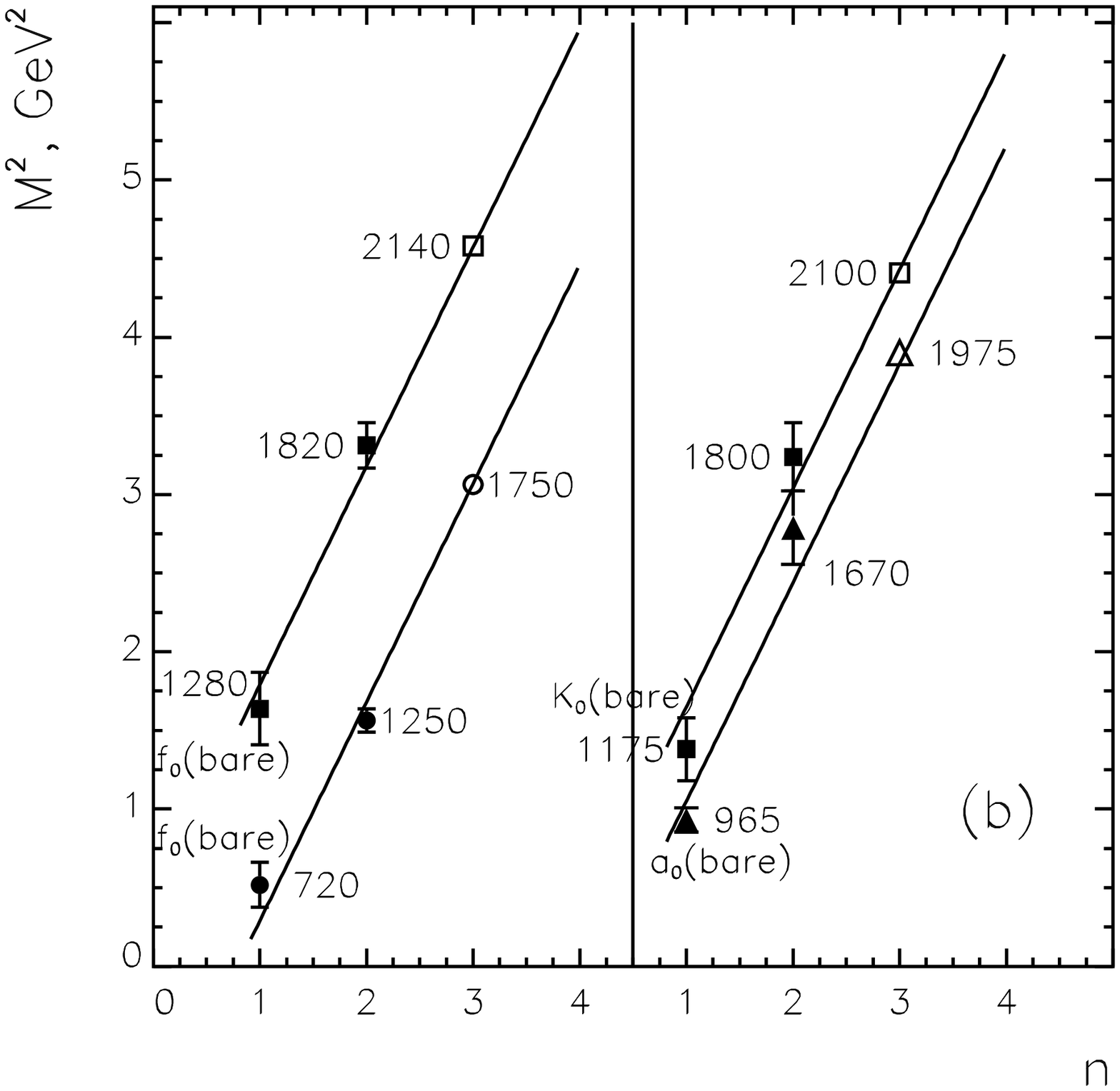,width=9cm}}
\caption\protect{Linear trajectories in $(n,M^2)$-plane for
scalar resonances (a) and scalar bare states (b).
Open points stand for predicted states.}
\end{figure}

\newpage
\begin{figure}
\centerline{\epsfig{file=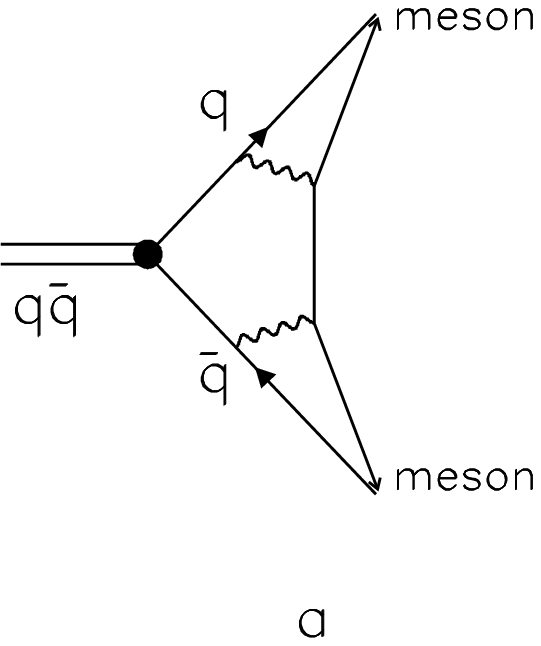,height=5cm}\hspace{0.5cm}
            \epsfig{file=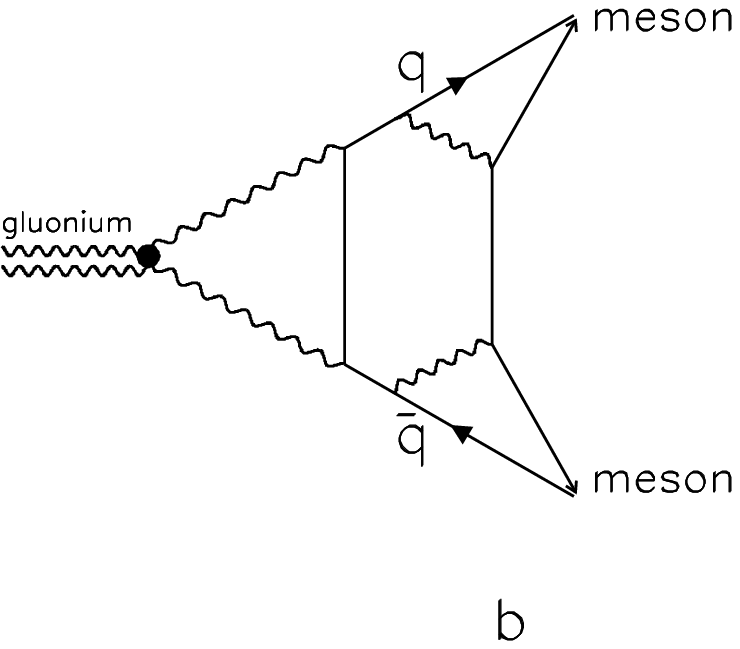,height=5cm}}
\vspace{0.5cm}
\centerline{\epsfig{file=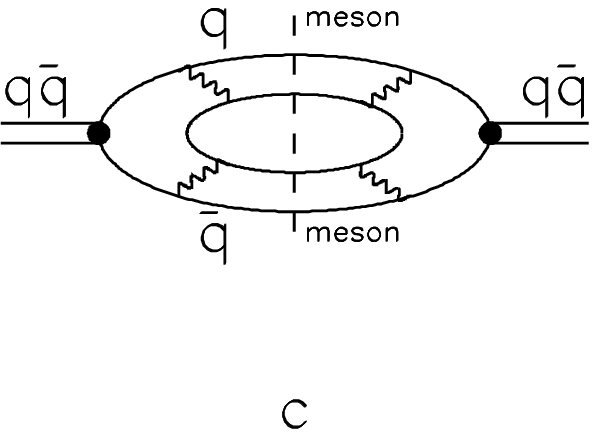,height=5cm}\hspace{0.5cm}
            \epsfig{file=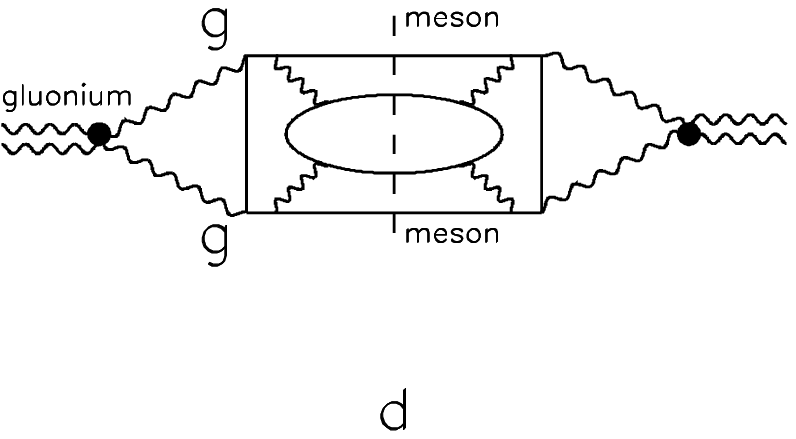,height=5cm}}
\vspace{-0.5cm}
\centerline{\epsfig{file=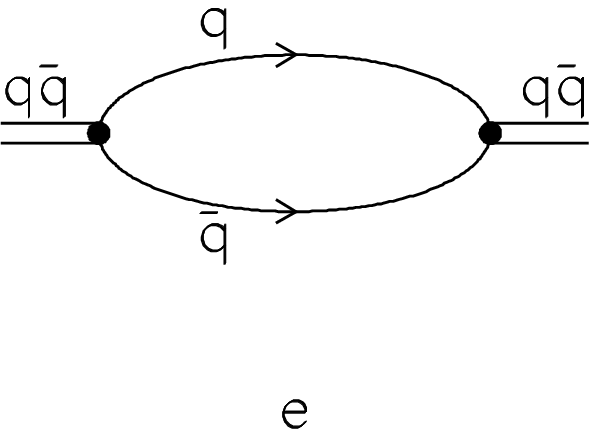,height=5cm}\hspace{0.5cm}
            \epsfig{file=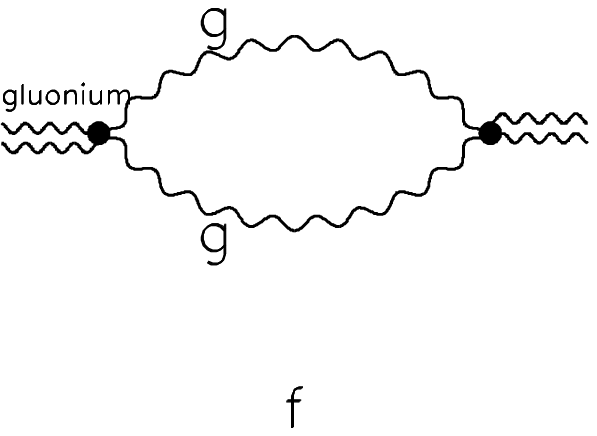,height=5cm}}
\vspace{-0.5cm}
\caption\protect{(a,b) Diagrams responsible for the transition of
scalar-isoscalar state into two pseudoscalar mesons, $f_0\to PP$,
in the leading $1/N_c$-expansion terms.
(c,d) Self-energy diagrams which determine $\sum g^2_a$
 (for quarkonium) and $\sum G^2_a$ (for gluonium); the cutting is shown
by dashed line.  (e,f) Self-energy diagrams which determine the order
of value of couplings for the transitions $quarkonium \to q\bar q$ and
$gluonium \to gg$.}
\end{figure}

\begin{figure}
\centerline{\epsfig{file=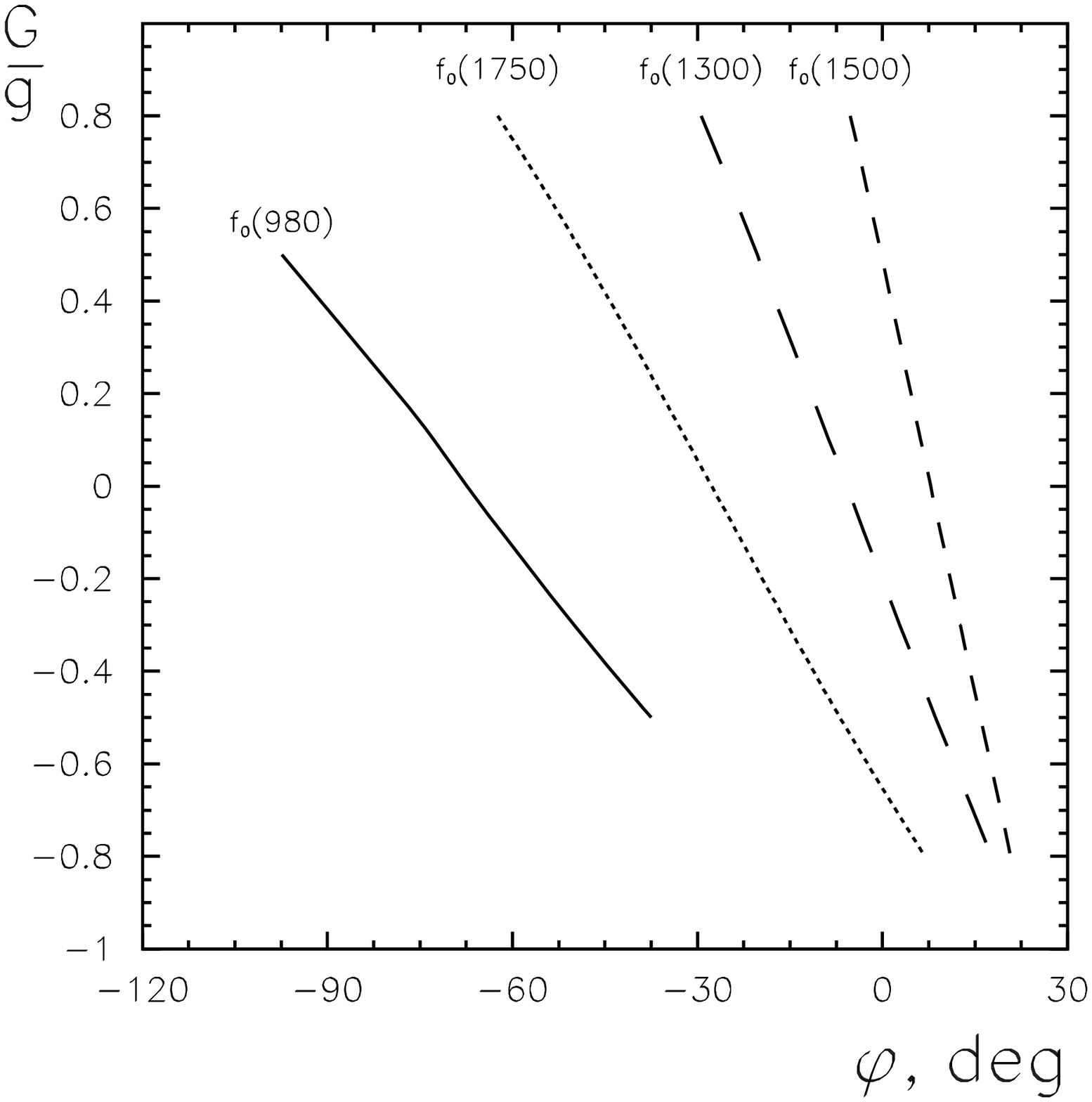,width=12cm}}
\caption\protect{$(G/g,\varphi)$-plot: the trajectories show correlations
between $G/g$ and $\varphi$ for
$f_0(980)$  (solid line),
$f_0(1300)$  (long-dashed line),
$f_0(1500)$ (dased line),
$f_0(1750)$  (dotted line).}
\end{figure}

\newpage
\begin{figure}
\centerline{\epsfig{file=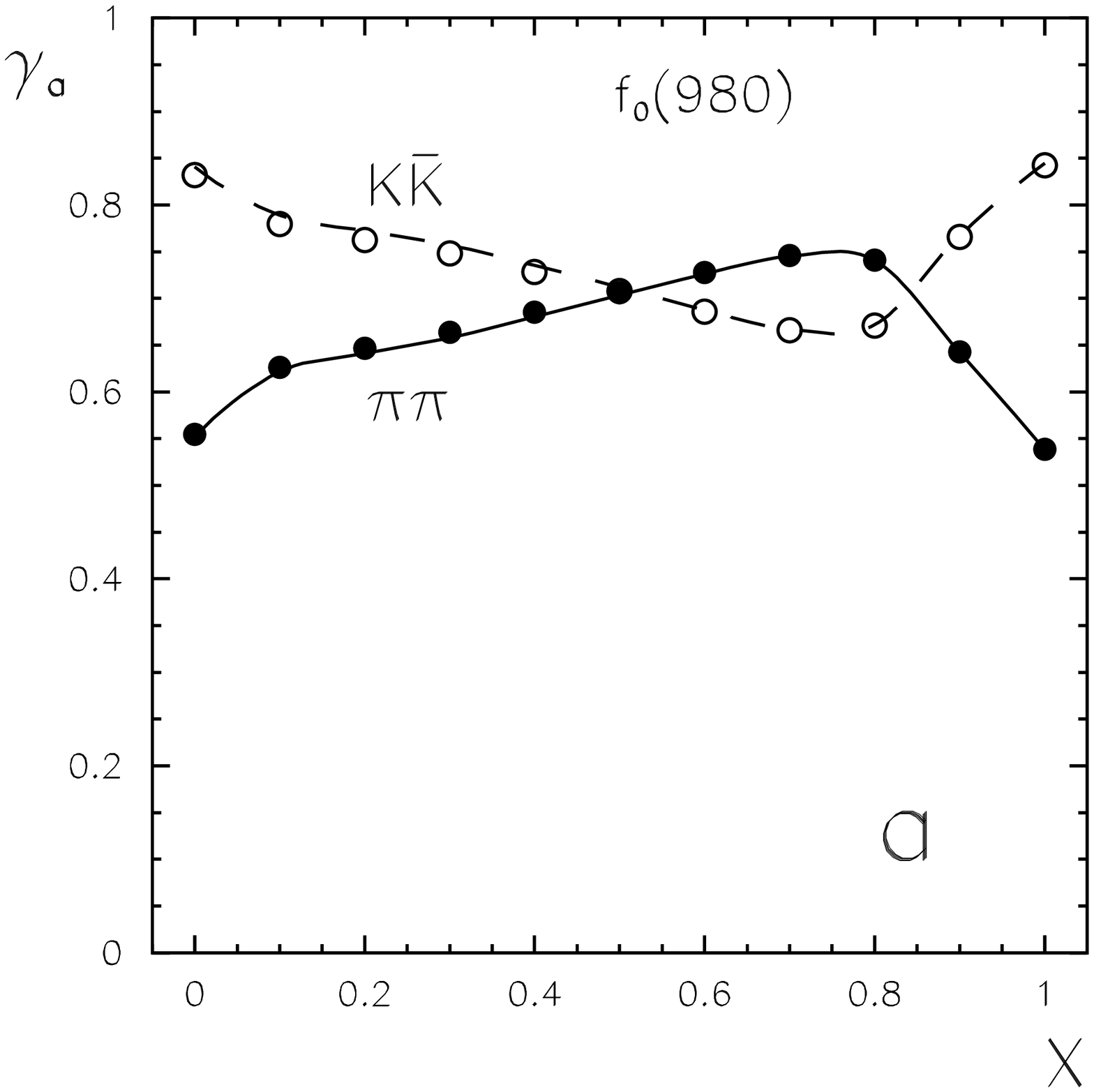,width=6cm}}
\centerline{\epsfig{file=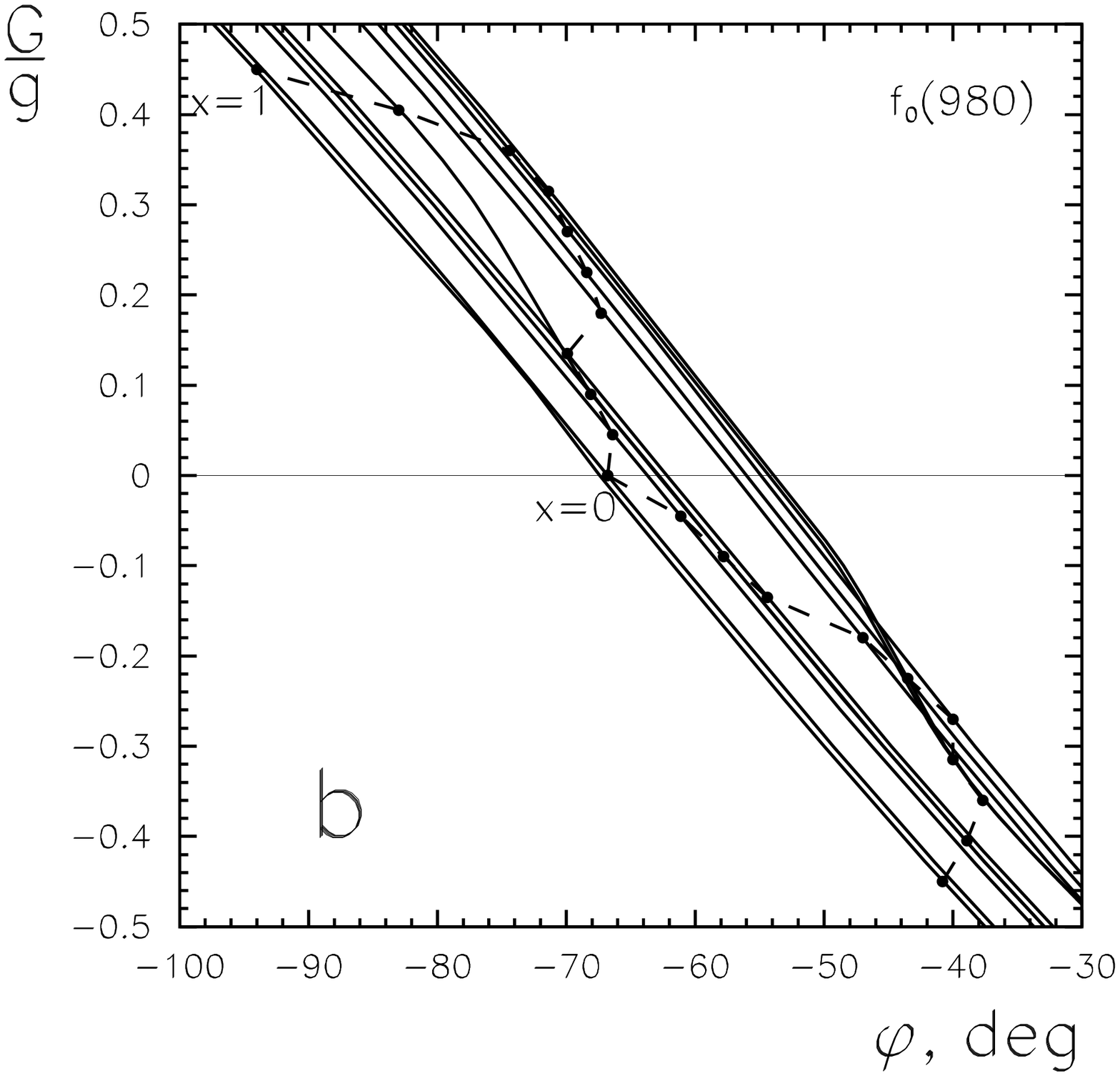,width=6cm}
            \epsfig{file=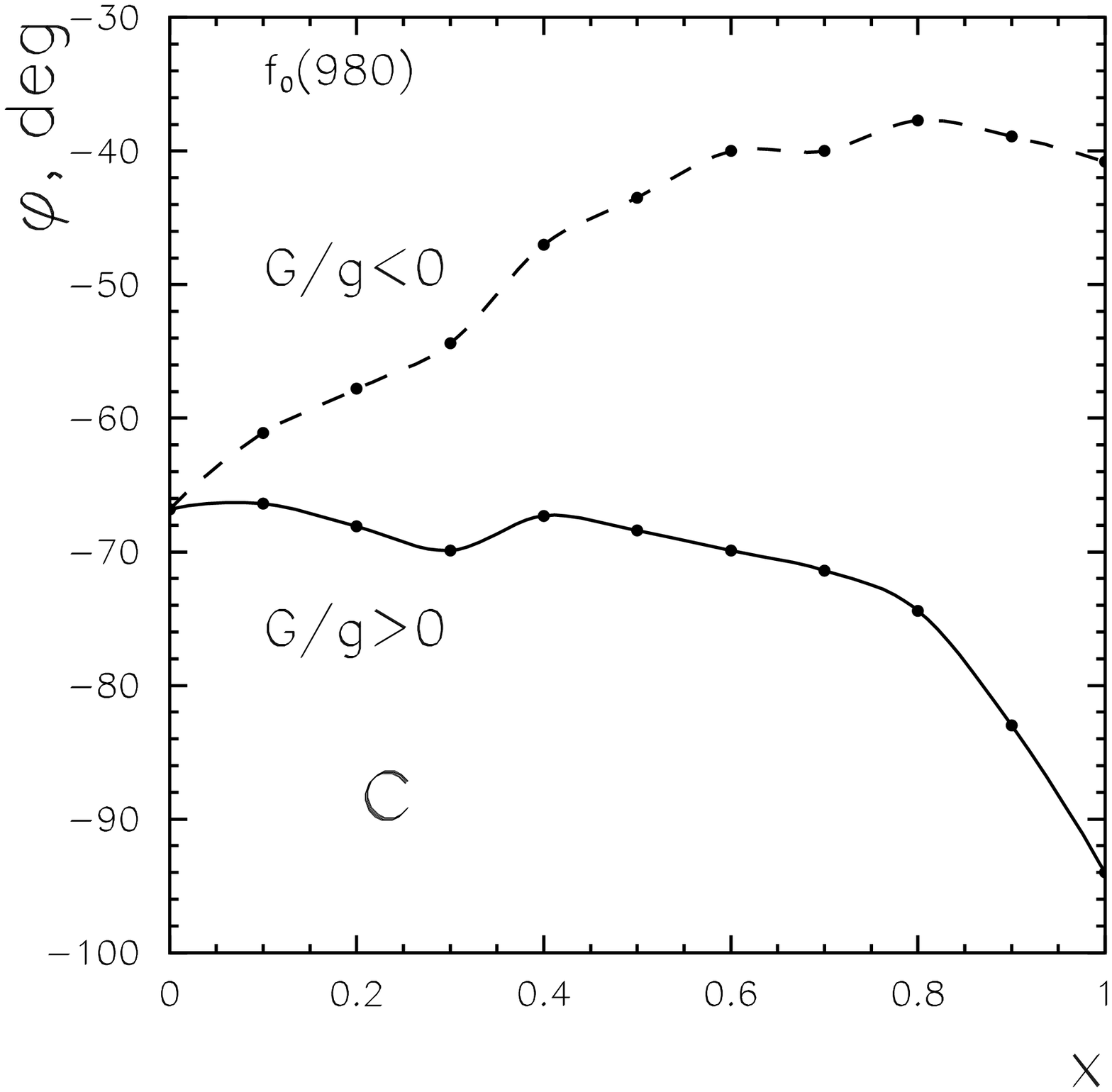,width=6cm}}
\caption\protect{The evolution of the $f_0(980)$ resonance parameters by
switching on/off the decay channels ($x=0$ corresponds to the bare state
while $x=1$ stands for the real meson). (a) Normalized coupling constants
$g_{\pi\pi}/\sqrt{g^2_{\pi\pi}+g^2_{K\bar K}}$ and
$g_{K\bar K}/\sqrt{g^2_{\pi\pi}+g^2_{K\bar K}}$ . (b) The
correlation curves
$(G/g,\varphi)$ at ten fixed values of $x$; dotted
curves stand for maximal accumulation of the gluonium component,
 $W_{gluonium}=20\%$. (c) The evolution of $\varphi$ in
the $q\bar q$ component $q\bar q =n\bar n\cos\varphi+s\bar
s\sin\varphi$ at maximal accumulation of the gluonium component: solid
curve stands for positive $G/g$ and dotted one to negative $G/g$.}
\end{figure}

\newpage
\begin{figure}
\centerline{\epsfig{file=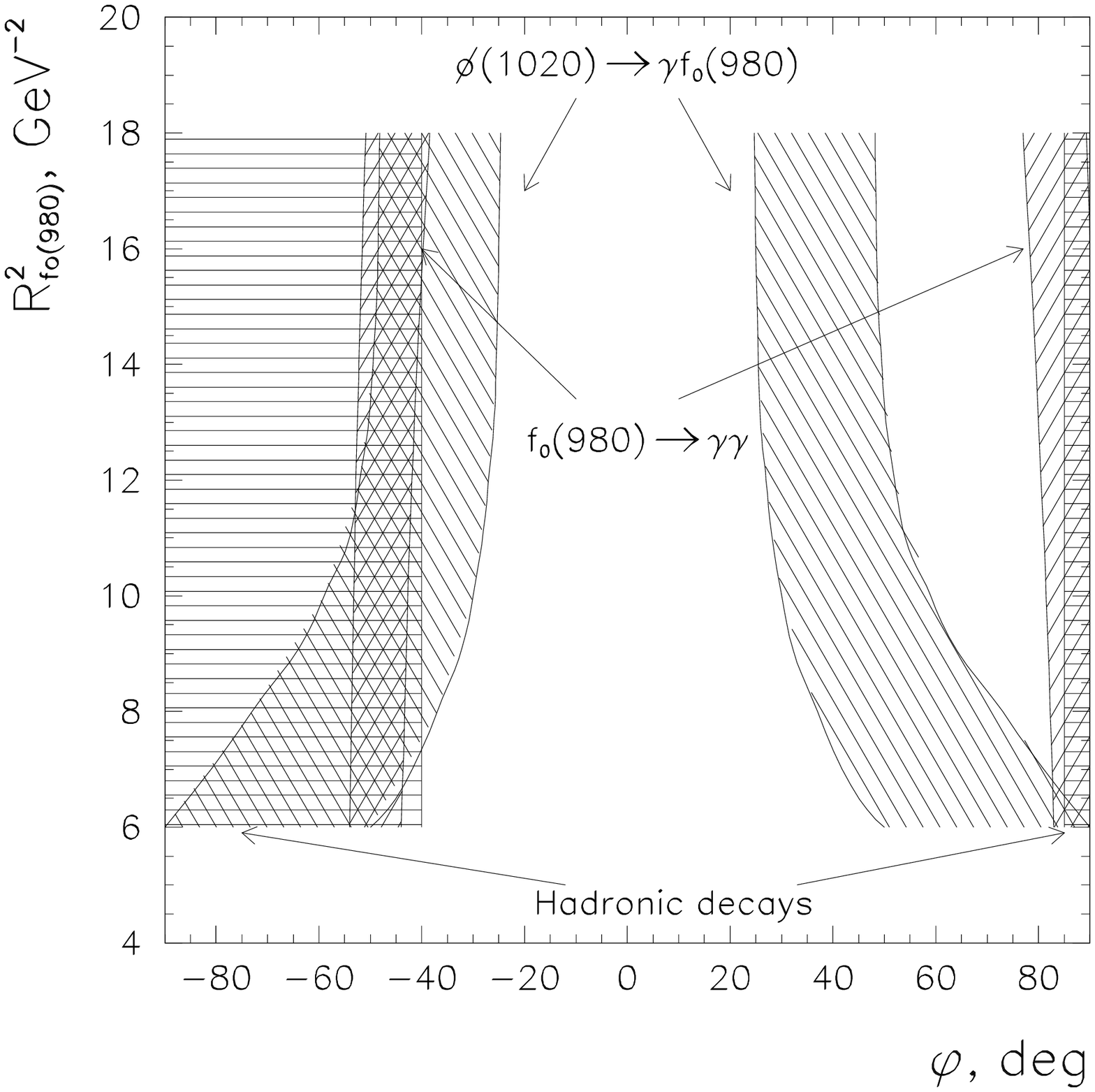,width=12cm}}
\caption\protect
{The $(\varphi,R^2_{f_0(980)})$-plot: the shaded areas are
the allowed ones for the reactions
$\phi(1020)\to\gamma f_0(980)$, $ f_0(980) \to \gamma\gamma$
and $ f_0(980) \to \pi\pi, K\bar K$.}
\end{figure}

\newpage
\begin{figure}
\centerline{\epsfig{file=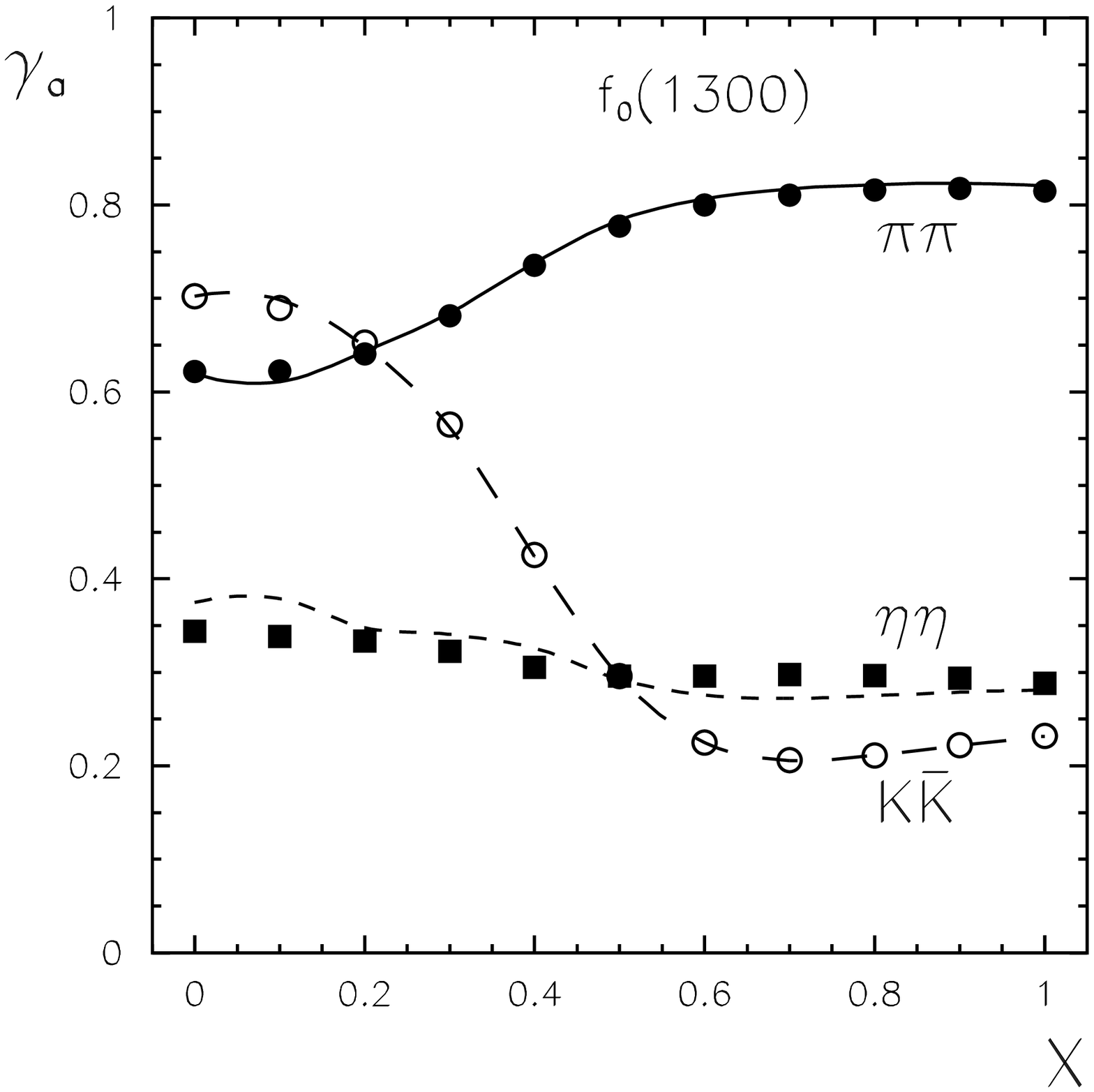,width=6cm}
            \epsfig{file=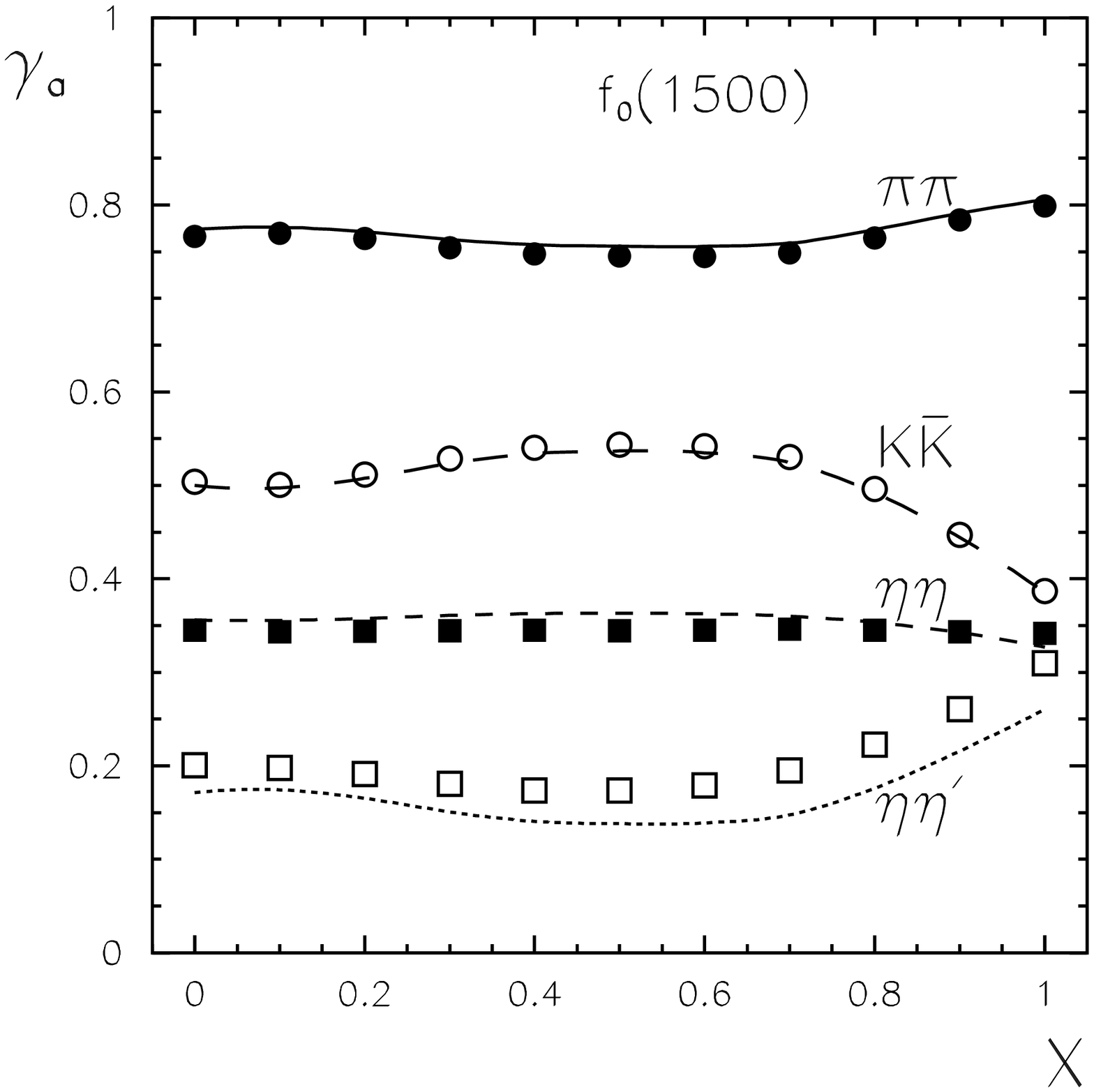,width=6cm}}
\centerline{\epsfig{file=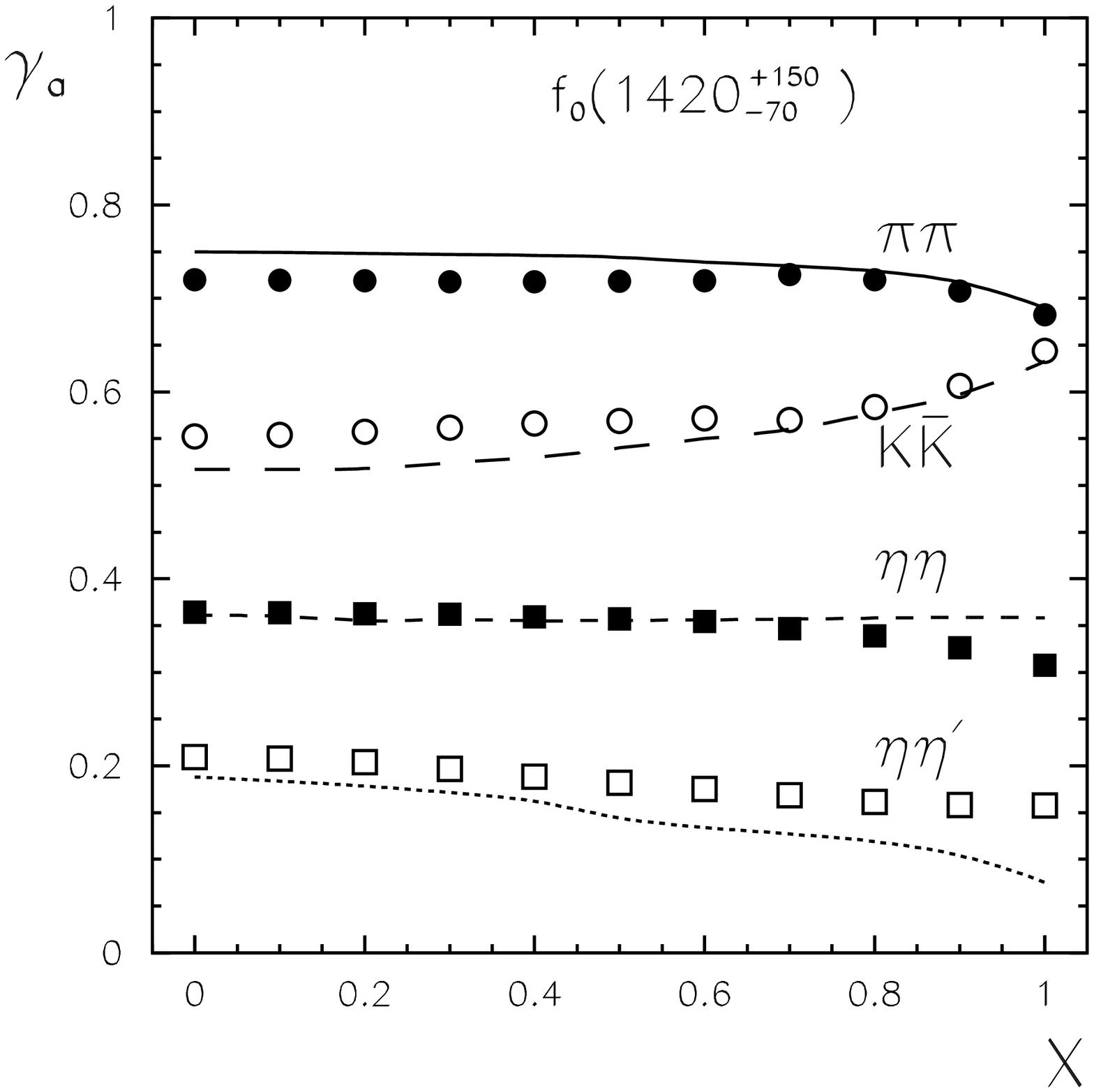,width=6cm}
            \epsfig{file=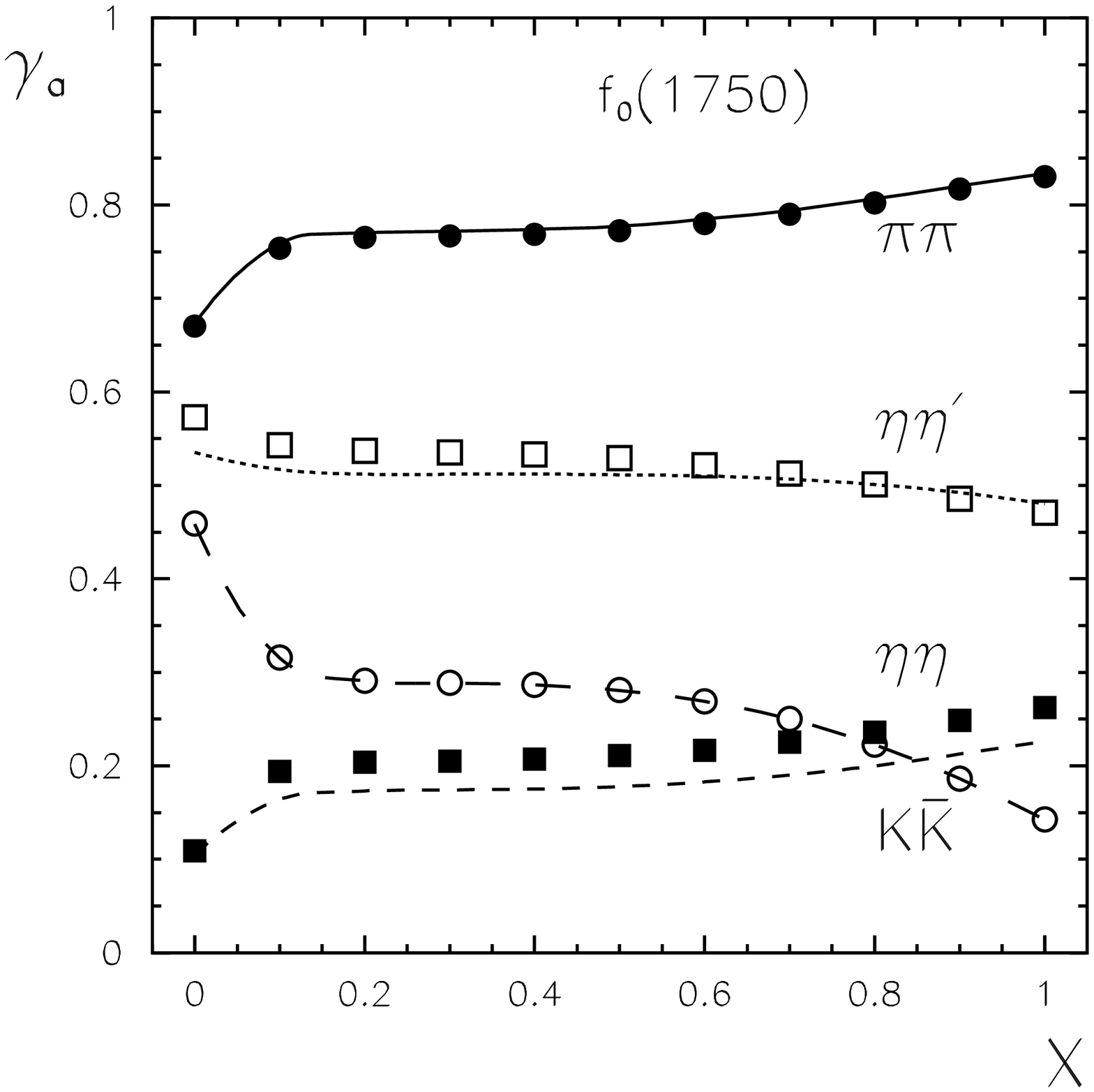,width=6cm}}
\caption\protect{ Evolution of the normalized couplings
$\gamma_a g_a/\sqrt{\sum_b g^2_b}$ within onset
of the decay channels for
$f_0(1300)$, $f_0(1500)$, $f_0(1420\; ^{+\;150}_{-70})$,
$f_0(1750)$. Curves demonstrate the description of
couplings by Eqs. (16). }
\end{figure}

\newpage
\begin{figure}
\centerline{\epsfig{file=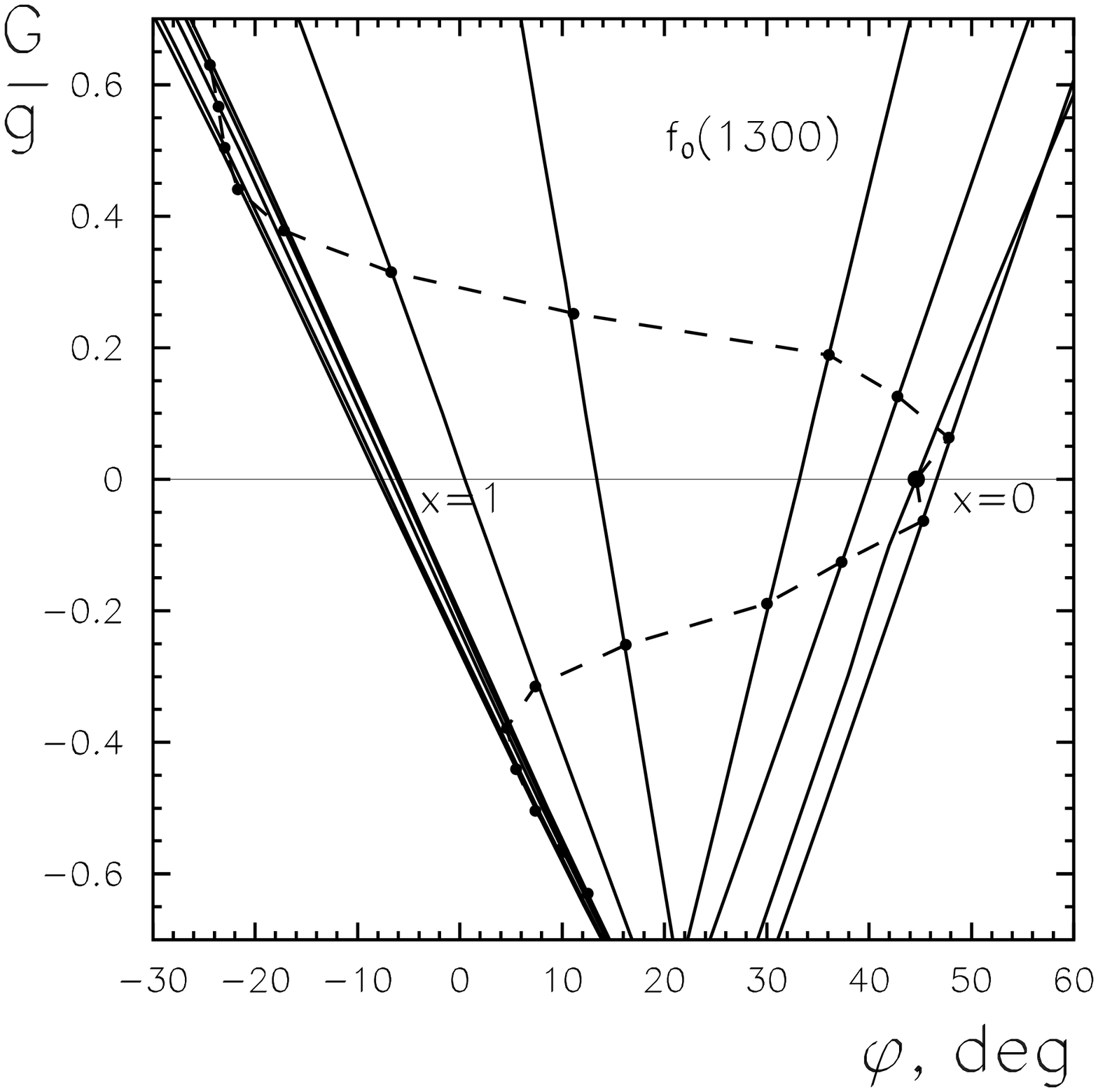,width=7cm}
            \epsfig{file=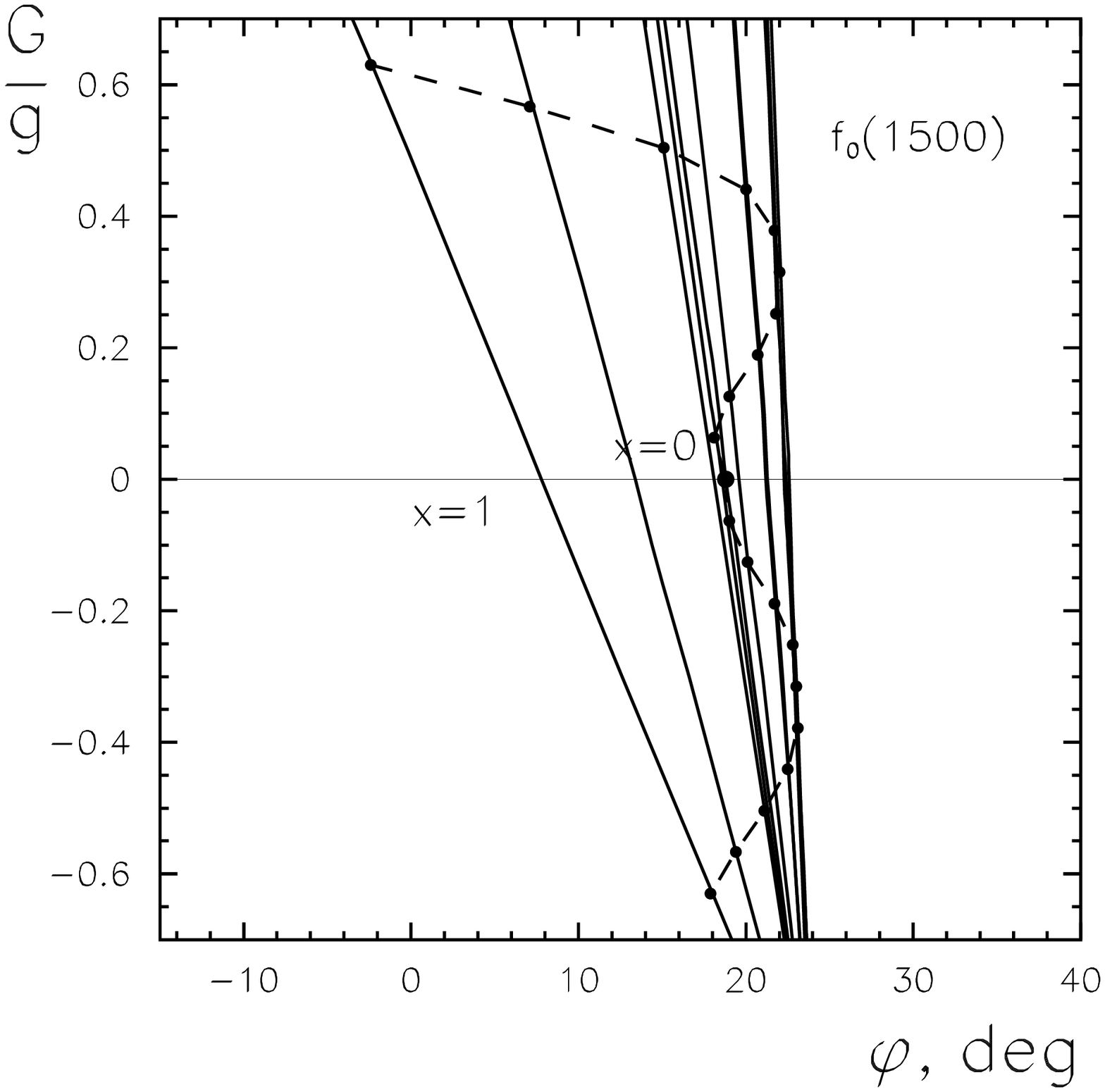,width=7cm}}
\centerline{\epsfig{file=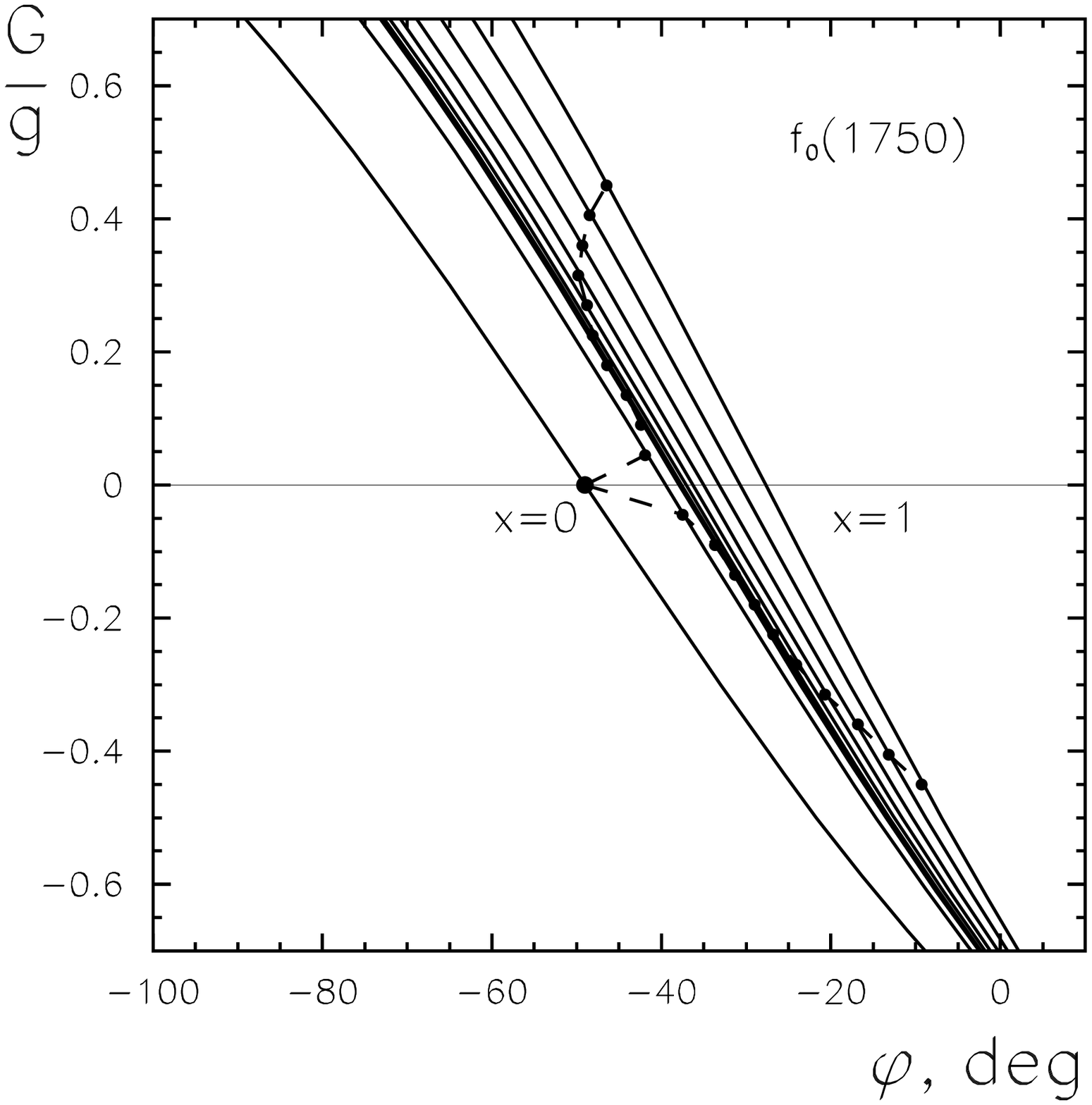,width=7cm}
            \epsfig{file=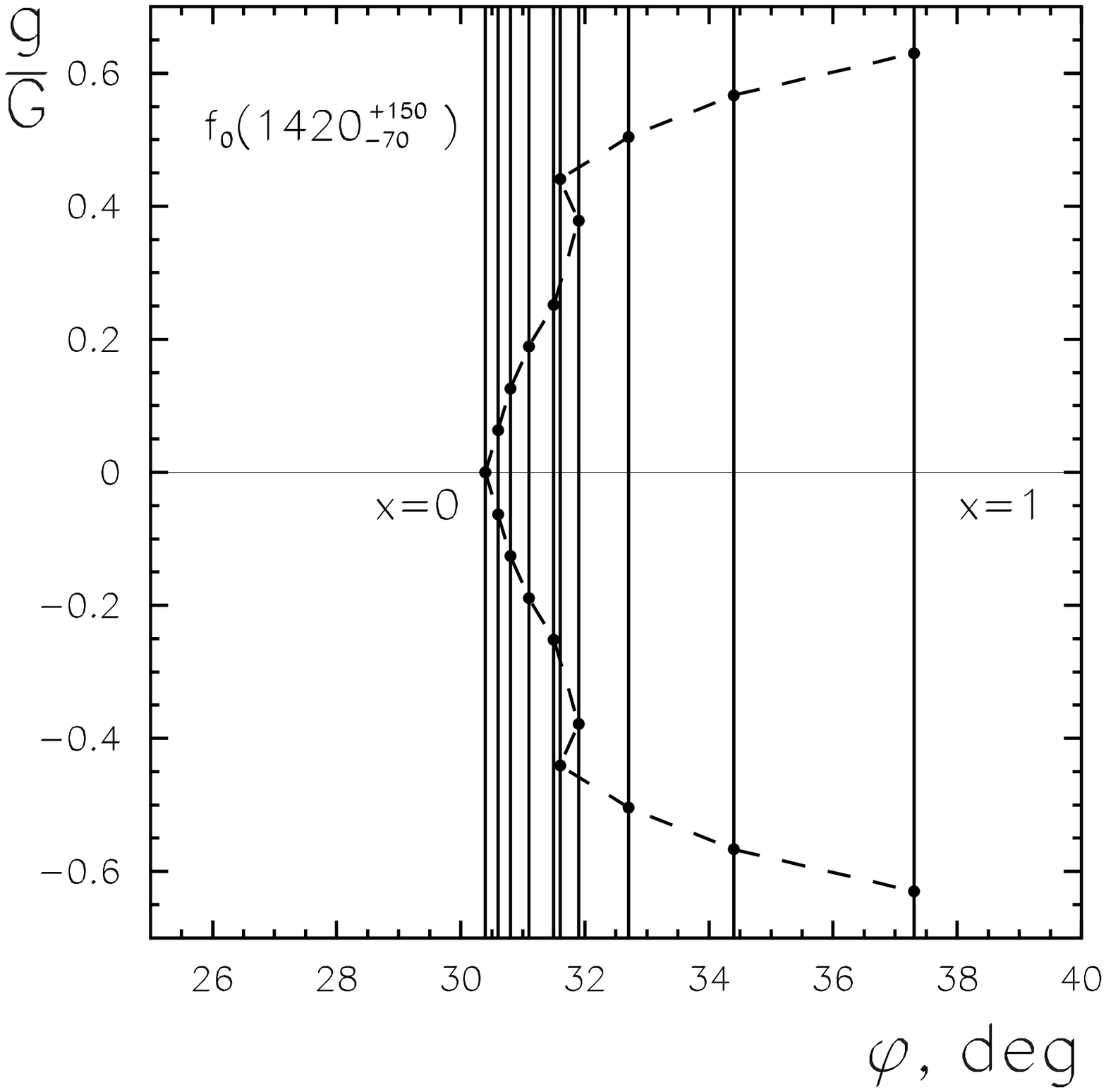,width=7cm}}
\caption\protect{Curves of correlations
$(G/g,\varphi)$ for
$f_0(1300)$, $f_0(1500)$,
$f_0(1750)$ and $(g/G,\varphi)$ for
$f_0(1420\; ^{+\;150}_{-70})$.
 at ten fixed values of $x$; dotted
curves stand for maximal accumulation of the gluonium component,
$W_{gluonium}=40\%$.  }
\end{figure}

\newpage
\begin{figure}
\centerline{\epsfig{file=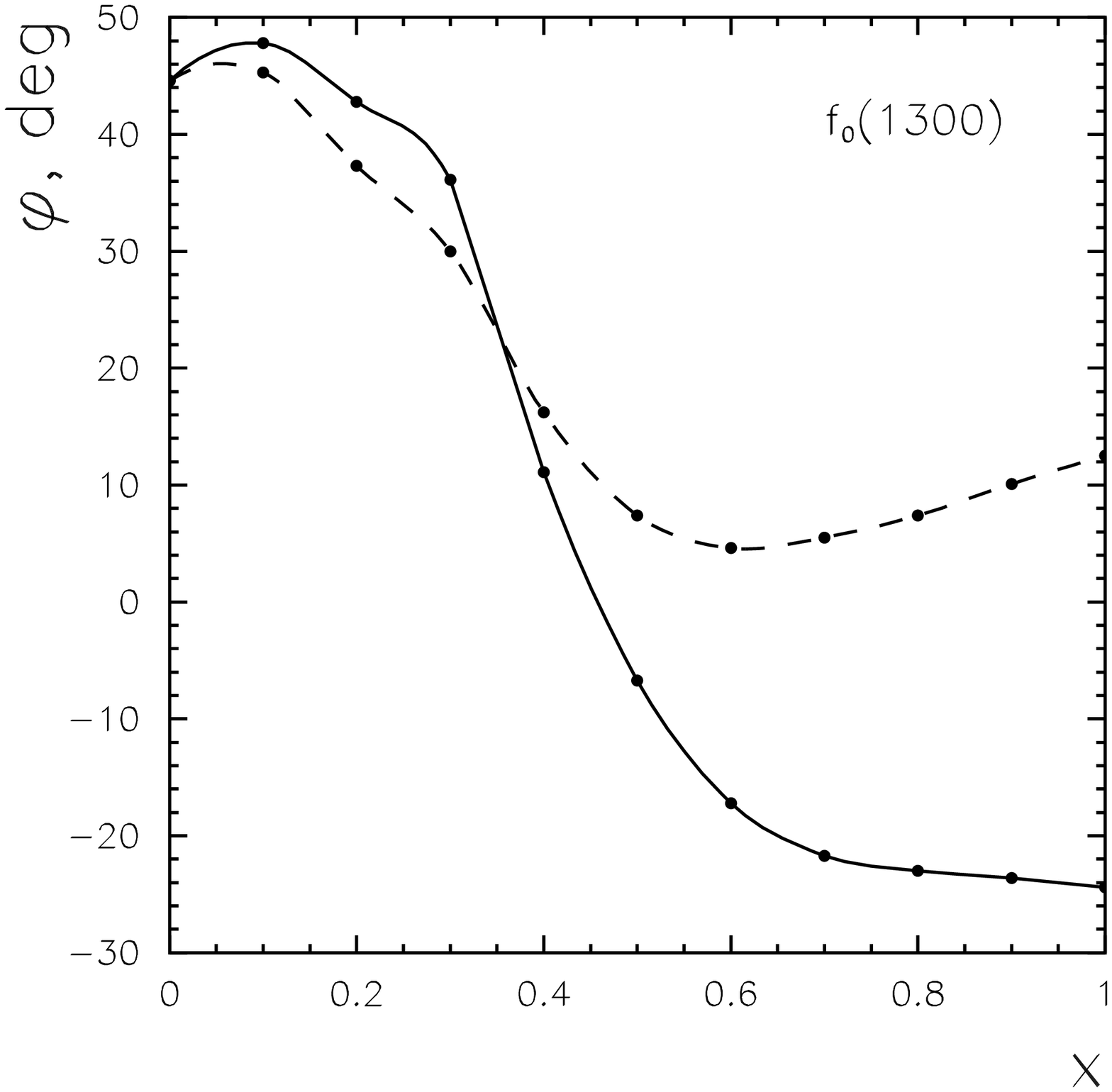,width=7cm}
            \epsfig{file=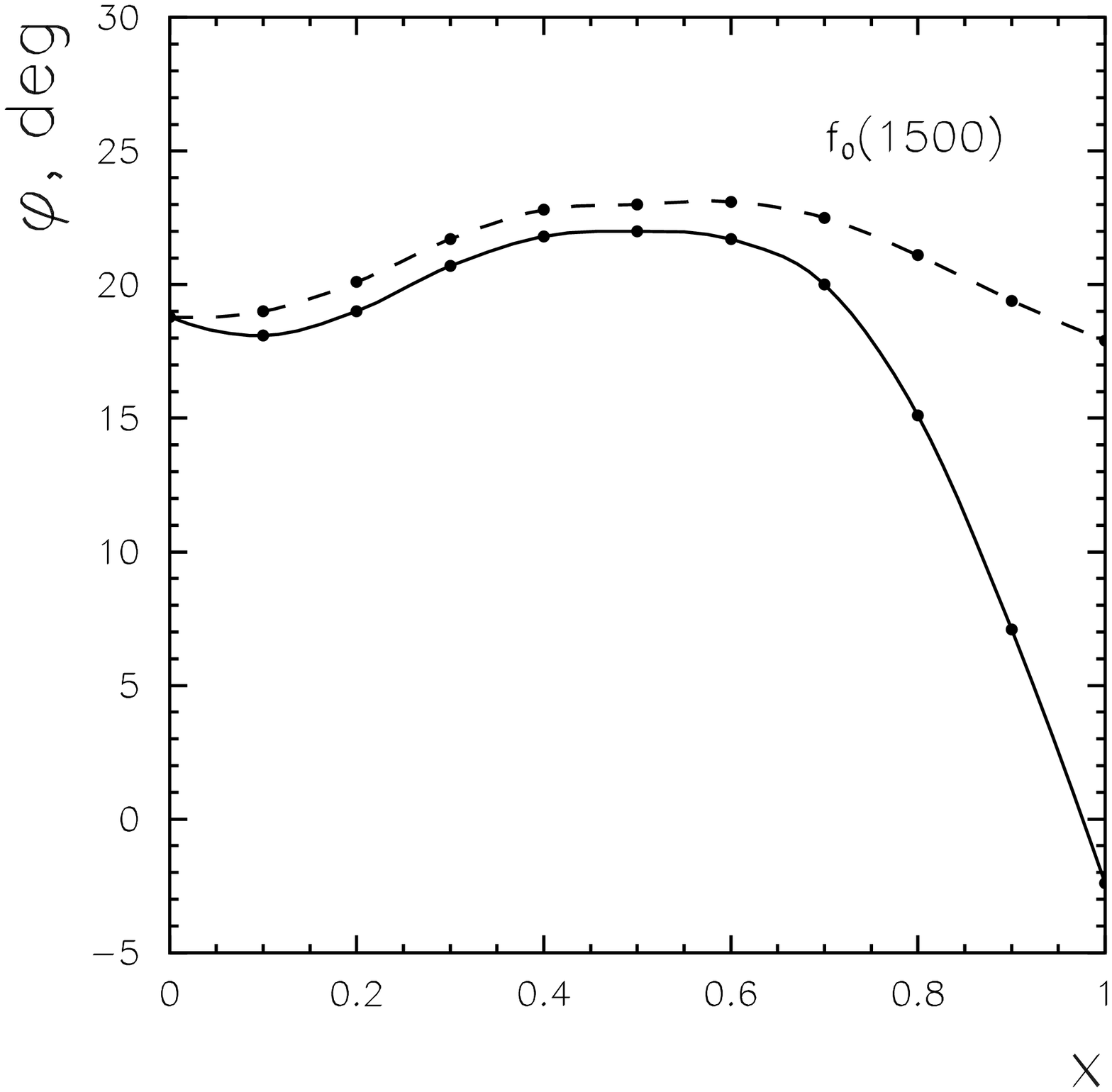,width=7cm}}
\centerline{\epsfig{file=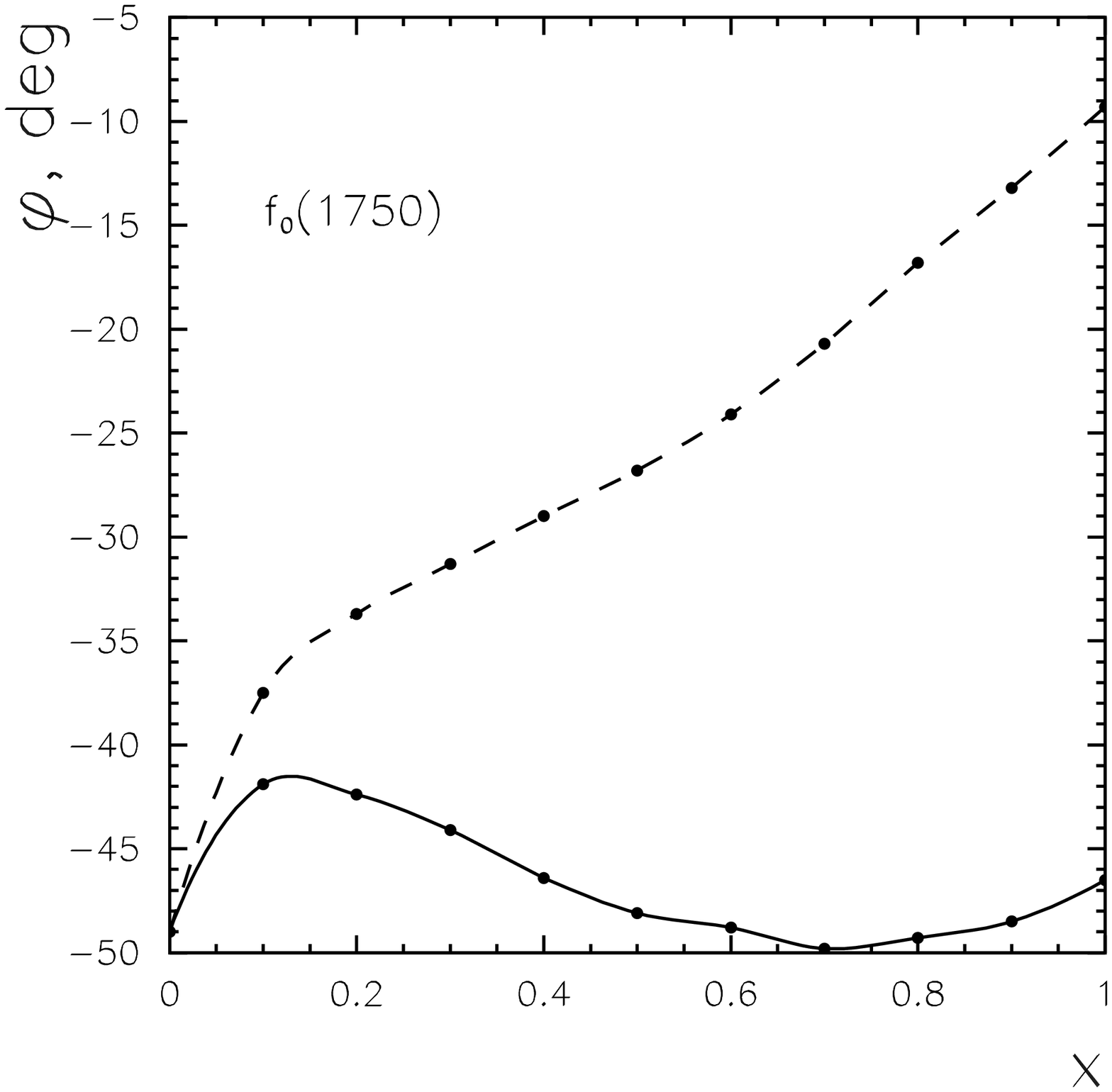,width=7cm}
            \epsfig{file=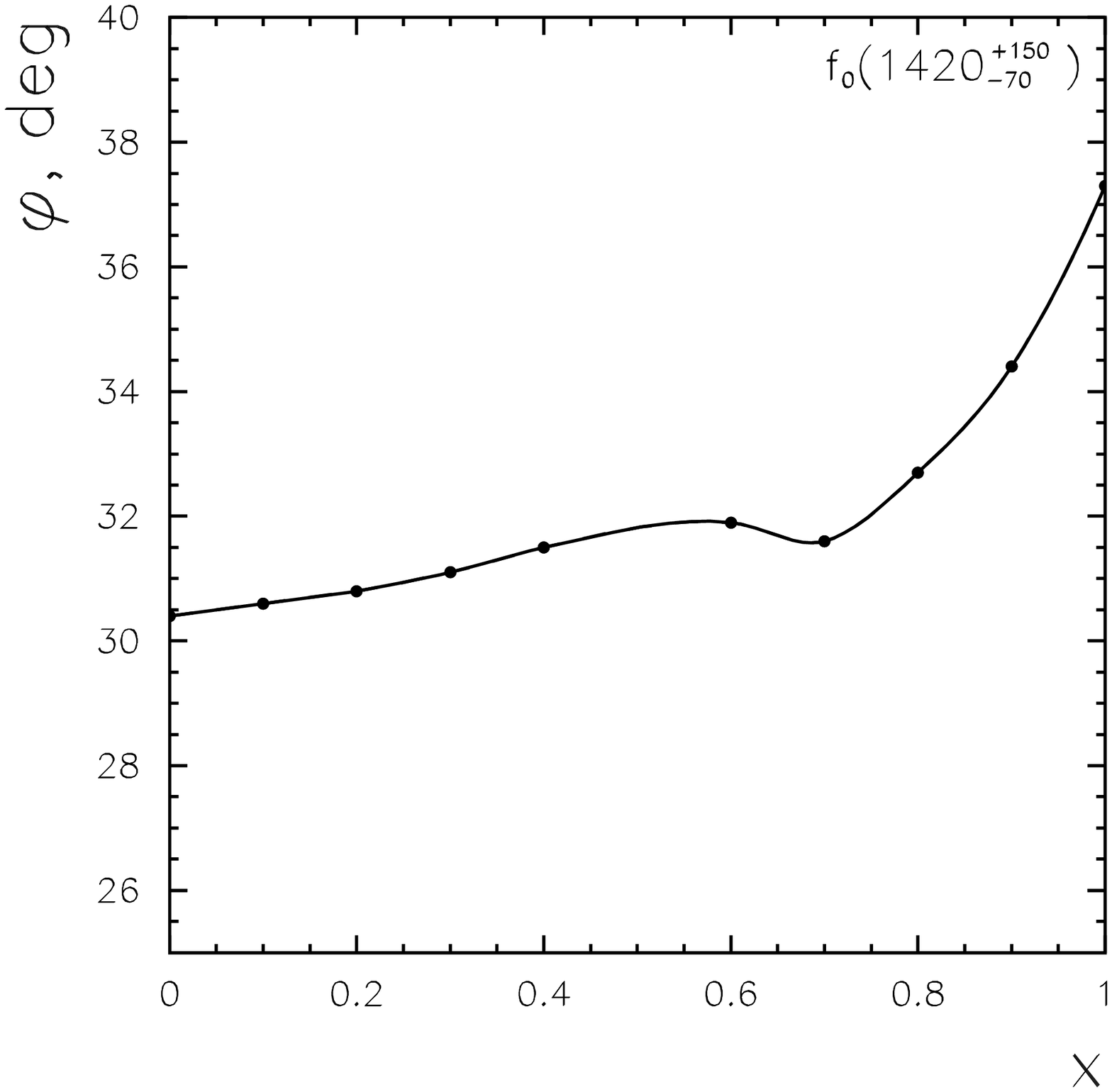,width=7cm}}
\caption\protect{
 The evolution of $\varphi$ in
the $q\bar q$ component $q\bar q =n\bar n\cos\varphi+s\bar
s\sin\varphi$ at maximal accumulation of the gluonium component,
$W_{gluonium}=40\%$; solid
curves stand for positive $G/g$ and dotted ones to negative $G/g$.
For $f_0(1420\; ^{+\;150}_{-70})$, the $\varphi$ does not depend on
the value of the adopted $q\bar q$ component. }
\end{figure}

\end{document}